\documentclass[preprint,showpacs,preprintnumbers,amsmath,amssymb]{revtex4}

\usepackage{graphicx}
\usepackage{dcolumn}
\usepackage{bm}

\begin{document}

\preprint{}

\title{
Semimicroscopic algebraic description of $\alpha$-clustering in $^{22}$Ne 
}

\author{G. L\'evai}
\email{levai@atomki.mta.hu}

\affiliation{%
Institute for Nuclear Research, Hungarian Academy of 
Sciences (MTA Atomki), H-4001 Debrecen, P.O. Box 51. Hungary
}%

\begin{abstract}
The alpha-cluster states of $^{22}$Ne are studied within the framework 
of the semimicroscopic algebraic cluster model (SACM). The band structure,
energy spectrum as well as E2 and E1 transitions are calculated and are 
compared with the experimental data. The results are also compared with 
those obtained from two microscopic models: the deformed-basis 
antisymmetrized molecular dynamics (DAMD) approach and the 
generator-coordinate method (GCM). It is found that the prominent bands 
obtained in the latter  frameworks all have equivalents in the SACM and 
the agreement between the calculated spectroscopic properties is rather 
good, especially for positive-parity states. 
\end{abstract}

\pacs{21.60.Gx, 21.60.Fw, 27.30.+t}

\keywords{}

\maketitle

\section{Introduction}

Clusterization is a special collective pheomenon that characterizes 
nuclei throughout the chart of nuclides. The most characteristic 
examples for clusterization have been found in light nuclei. These 
configurations show a variety both in the type of clusters (ranging 
individual nucleons to $^{12}$C and $^{16}$O) and their number 
(from two- to four-body systems and even beyond). The most typical 
cluster is the $\alpha$ particle, which is a tightly bound system 
of two protons and two neutrons. $\alpha$ cluster states 
are present in the low-lying spectrum of many light nuclei throughout the 
$p$ and $sd$ shell and typically appear as the members of well  
deformed collective bands. 

In the theoretical description of clustering it is essential to 
include the relative motion of the clusters as the most important 
degree of freedom. However, since the cluster system as a whole consists 
of interacting fermions (nucleons), the effects of the Pauli exclusion 
principle also have to be taken into account. This can be done by 
using a basis in which the single-nucleon states appear in a fully 
antisymmetrized form. Such basis is used in {\it microscopic} cluster 
models \cite{micr_cl_m}, which also employ microscopic nucleon-nucleon 
interactions. {\it Phenomenological} cluster models follow a technically 
simpler approach in that their model space does not observe strictly  
the Pauli principle, and they employ effective cluster-cluster 
interactions. This is the case, for example, with some potential models 
\cite{buck}. {\it Semimicroscopic} cluster models typically combine 
a microscopic model space with effective interactions. Examples for 
this are the orthogonality condition model (OCM) \cite{ocm} and the 
semimicroscopic algebraic cluster model (SACM) \cite{sacm}. 

The SACM makes heavy use of the SU(3) algebra, which appears in 
the description of both the relative motion and that of the internal 
cluster structure, where it accounts for the orbital sector in 
terms of Elliott's SU(3) model \cite{elliott,harvey}. The SU(3) algebra is 
the symmetry algebra of the three-dimensional harmonic oscillator and 
it is also used in the construction of the harmonic oscillator (or SU(3)) 
shell model. The connection between the shell and cluster approaches 
was established more than fifty years ago in the harmonic oscillator 
approximation \cite{wildermuth} making use also of the SU(3) formalism 
\cite{bayman}. Based on this connection the model space of the SACM is 
a subset of the fully antisymmetrized SU(3) shell model space, which 
also takes into account the assumed cluster configuration. It contains 
states with good SU(3) symmetry, which are assigned to SU(3) irreducible 
representations (irreps). This symmetry-dictated truncation of the SU(3) 
shell model space typically contains the most deformed states of the joint 
nucleus, reflecting the cluster configuration of the system, but less 
deformed states are also included usually. In fact, this method seems 
to be fairly effective in studying the shape isomers (e.g. superdeformed 
and hyperdeformed states) and the clusterizations associated with them 
\cite{csj}. 

It is notable that the SU(3) symmetry seems to be a good approximation 
for many nuclei in the $p$ and the $sd$ shell apart from the cluster 
picture too. In Ref. \cite{ijmpe} it was shown that an extremely 
simplified, essentially parameter-free Hamiltonian that contains only 
the harmonic term and the second-order Casimir operator of the SU(3) 
algebra is able to account in this region for the trends of some basic  
observables (e.g. energy of the first excited opposite-parity level) 
of nuclei. 

More recently much attention has been paid to 
nuclei in which a core system consisting of closed-shell clusters 
is surrounded by a few neutrons. In these systems the neutrons can be 
pictured as moving on molecular orbitals around the core configuration. 
Typical examples are the Be isotopes, in which the neutrons surround 
two $\alpha$ particles \cite{be+xn}. Neon isotopes have been proposed 
as the next possible examples, in which the neutrons orbit around the 
$^{16}{\rm O}+\alpha$ core \cite{kim9}. These systems can also be 
described in the antisymmetrized molecular dynamics (AMD) approach 
\cite{kim6-8} or its extension, the deformed-basis AMD (DAMD) 
\cite{kim12}. This microscopic approach does not assume a 
cluster structure from the beginning and can describe systems in which 
both the shell and cluster structures, or their mixture play a role. 

Here we discuss the $^{22}$Ne nucleus in terms of the semimicroscopic 
algebraic cluster model. Our aim is not only the theoretical interpretation 
of the experimental spectroscopic information (band structure, spectrum, 
electromagnetic transitions) but also comparing the performance of the 
SACM with that of microscopic approaches, the GCM \cite{desc88,desc03} 
and the DAMD \cite{kimura}. The SACM shares some common features with 
these models and it is worthwhile to investigate whether these manifest 
themselves in the results. The GCM employs a microscopic model space 
constructed from harmonic oscillator orbitals such that the oscillator 
parameter is common for each cluster. This technical assumption is also 
used in defining the model space of the SACM. Furthermore, both models 
consider excited states of the core cluster. In the SACM these states 
are those belonging to the ground-state SU(3) configuration of the core, 
while in the GCM they can be selected independently. The common feature 
of the SACM and the DAMD is the ability of these models to give a joint 
description of shell-like and cluster-like configurations in the same 
system. In the SACM this is again due to the SU(3) basis, which is 
a subset of the SU(3) shell model basis. As such, it contains states 
with well developed cluster structure typically appearing around the 
threshold energy of the assumed cluster configuration, but also 
shell model-like states in or around the ground-state region. 

The paper is structured as follows. 
In Sec. \ref{sacm_intro} we present the main features of the 
SACM. Section \ref{22ne_appl} contains the SACM description of the 
$^{22}$Ne nucleus in terms of the $^{18}$O+$\alpha$ configuration, while 
in Sec. \ref{compare} the results are compared with those obtained 
from the GCM and DAMD approaches. Finally, conclusions are drawn 
in Sec. \ref{concl}.

\section{The Semimicroscopic Algebraic Cluster Model}
\label{sacm_intro}

Here we review the basics of the SACM \cite{sacm} for the 
core+$\alpha$ states of even-even nuclei. This model has been applied 
previously to a number of core+$\alpha$ systems \cite{prc92,cunda}. 
It was shown that the parameters fitted to the spectrum of nuclei in 
the $A=16$ to 20 domain vary in a consistent way \cite{plb96}.  
This correlated behavior was interpreted for $A=18$ to 20 nuclei in 
terms of a new cluster supersymmetry scheme \cite{jpg07}, in which the 
bosonic degree of freedom was associated to the excitations of the 
relative motion, while the fermionic structure was represented by 
0, 1 or 2 holes in the $p$ shell. Other cluster structures with 
heavier clusters were also discussed in terms of the SACM \cite{other}.

\subsection{The model space}
\label{two}

The relative motion (indexed by `$R$') is described by the group structure 
\begin{align}
&\hspace{0.3cm}U_R(4) \hspace{0.3cm} \supset \hspace{0.3cm} SU_R(3) 
\hspace{0.3cm} \supset \hspace{0.3cm} SO_R(3) \hspace{0.3cm} \supset
\hspace{0.3cm} SO_R(2)
\nonumber \\
&\lbrack N,0,0,0 \rbrack  \hspace{0.85cm}(n_{\pi} , 0 )   
\hspace{1.5cm}L_R  \hspace{1.7cm}M_R ,
\label{u3}
\end{align}
where
$n_{\pi} = N, N-1,...,1,\ 0$,
$L_R =  n_{\pi}, n_{\pi} -2,...,\ 1 ~ \text{or} ~ 0$, 
$M_R =  L_R, L_R-1, ...,\ -L_R$.
Here $N=n_{\pi}+n_{\sigma}$ is the total number of bosons, $n_{\pi}$ 
is the number of dipole bosons, i.e. oscillator quanta assigned 
to the relative motion of the clusters, while $n_{\sigma}$ is the 
number of scalar bosons. The role of these latter bosons is 
essentially introducing an upper cut-off in the model space, making 
it finite. There is also a lower cut-off in the number of dipole 
bosons due to the Wildermuth condition, which is necessary to 
take into account the Pauli principle. This minimal number of 
$n_{\pi}$ is determined by the structure of the two clusters. 
In core+$\alpha$  systems it is determined by the core nucleus. 
For cores taken from the $sd$ shell it is usually $n_{\pi\ {\rm min}}=8$ 
corresponding to lifting the four nucleons of the $\alpha$ particle 
to the $2\hbar\omega$ shell from the $0\hbar\omega$ shell. 
The parity assigned to the relative motion is $(-1)^{n_{\pi}}$. 

It is important to note that the procedure that determines the 
model space by matching the SU(3) cluster model space with the 
SU(3) shell model space eliminates the Pauli-forbidden states, 
so it also enforces the Wildermuth condition. 

The core nucleus is described in terms of Elliott's SU(3) shell 
model 
\cite{elliott,harvey}. 
 This model assigns an SU$_C$(3) representation to the 
orbital structure of the core nucleus (hence the index $C$), while 
the spin-isospin structure is described by Wigner's SU(4) 
supermultiplet scheme that handles the spin and isospin structure 
in a combined way. 
The orbital and isospin structures are interrelated by the 
antisymmetry requirement. In practice this means that in the 
maximally symmetric orbital structure that characterizes the 
ground-state configuration of the core nucleus comes together 
with maximally antisymmetric spin-isospin structure, which, 
in general, prescribes low values of $S$ and $T$. For even-even 
nuclei this implies $S=0$ and $T=(N-Z)/2$. This is the case for 
the $\alpha$ particle too, with $S=0$ and $T=0$. In this case, 
furthermore, the ground-state orbital structure is described by 
the scalar (0,0) SU(3) representation with $L=0$ orbital angular 
momentum. In practice this means that the internal structure of 
the $\alpha$ particle does not have direct influence on 
the construction of the model space. 

It is worth mentioning that it is possible to consider 
in the core sector configurations that correspond to excited 
states either with $0\hbar\omega$ or even $1\hbar\omega$. 
However, this is not typical and has not been applied up to now. 

The orbital part of the core+$\alpha$ system is determined by 
the SU(3) coupling of the representations assigned to the 
relative motion and the core nucleus. In the general case this is done 
as  
\begin{align}
&& SU_{C}(3) \otimes SU_R(3) \hspace{0.3cm} \supset 
  SU(3) \hspace{0.3cm} \supset \hspace{0.3cm} SO(3)
\hspace{0.3cm} \supset \hspace{0.3cm} SO(2)
\nonumber \\
&& (\lambda_C,\mu_C) \hspace{0.3cm} (n_\pi,0) \hspace{1.0cm}
  (\lambda , \mu )  \hspace{1.7cm} \kappa L \hspace{1.7cm} M\ .
\label{chain}
\end{align}
The $(\lambda,\mu)$ SU(3) irreps in (\ref{chain}) are 
obtained from the direct product 
\begin{equation}
\left( \lambda_C,\mu_C\right) \times \left(n_\pi ,0\right) 
= \sum_{\lambda\mu} m_{\lambda\mu}\left(\lambda , \mu \right)
~~~,
\label{anti}
\end{equation}
with multiplicity
$m_{\lambda\mu}$. When one of the SU(3) representations 
has either $\lambda=0$ or $\mu=0$ (as is the case for any 
core+$\alpha$ system), each resulting 
representation has multiplicity one. 

The cluster model space 
constructed in this way contains states (SU(3) representations) 
that are not allowed by the Pauli principle. The final model space 
is obtained after matching the set of SU(3) states from Eq. 
(\ref{anti}) with the fully antisymmetrized SU(3) shell model space. 
This procedure also eliminates the spurious center of mass excitations. 

The $(\lambda,\mu)$ SU(3) representations determine the SO(3) 
representations contained in them, i.e. 
the orbital angular momenta $L$. Here the $\kappa$ quantum number 
also plays an important role in resolving the multiplicities 
of the $L$ quantum numbers contained in the same SU(3) representation. 
The general rule is 
\begin{eqnarray}
\kappa={\rm min}[\lambda,\mu],\ {\rm min}[\lambda,\mu]-2,\dots 
1\ {\rm or}\ 0;
\nonumber\\  
L=\kappa, \kappa+1\dots \kappa+{\rm max}[\lambda,\mu]\ 
{\rm for} \ \kappa\ne 0\ {\rm and} 
\nonumber\\
L={\rm max}[\lambda,\mu]\ , {\rm max}[\lambda,\mu]-2\ , \dots 
1\ {\rm or}\ 0\ {\rm for} \ \kappa=0\ . 
\label{su3toso3}
\end{eqnarray}
In practice $\kappa$ identifies complete bands for even-even nuclei 
as, in general, it plays the role of $K$, the projection of the 
angular momentum on the symmetry axis of the nucleus. This means that 
each SU(3) representation contains one or more complete rotational 
bands in this case. Breaking this result down to the various SU(3) 
representations one finds that representations 
of the type $(\lambda,0)$ and $(\lambda,1)$ contain a single 
rotational band with $\kappa=0$ and 1, respectively. Representations 
of the type $(\lambda,2)$ contain two bands with $\kappa=0$ and 2, 
$(\lambda,3)$ also contain two bands with $\kappa=1$ and 3, while 
$(\lambda,4)$ contains three, with $\kappa=0$, 2 and 4. 

The SU(3) representations are also indicative about the deformation 
of the joint nucleus, because in the SU(3) scheme these quantum 
numbers are related to the difference of the distribution of 
oscillator quanta in the three directions. Accepting the usual 
notation and assuming that these numbers are related by 
$n_z\ge n_x\ge n_y$, the $\lambda$ and $\mu$ quantum numbers 
are expressed as $\lambda=n_z-n_x$ and $\mu=n_x-n_y$. This means 
that $\mu=0$ corresponds to a prolate deformation, while 
$\lambda=0$ to an oblate deformation. When neither $\lambda$ and 
$\mu$ are zero, then the nucleus is triaxial. Nevertheless, small 
$\mu$ and large $\lambda$ describes a near prolate structure  
\cite{harvey}. 

The parity of the $\alpha$-cluster states is determined by $n_{\pi}$, 
because the parity assigned to the core configuration is 
uniquely defined, as is the parity of the $\alpha$ paricle. 
The general form of the SACM basis for an even-even core+$\alpha$ 
configuration is written as
\begin{equation}
\vert N n_{\pi}, (\lambda_C,\mu_C); (\lambda,\mu)\kappa L T\rangle\ .
\label{basis}
\end{equation}
As it was discussed above, this basis is fully microscopic by 
construction. 
Note that the short-hand notation $n_{\pi}(\lambda,\mu)\kappa$ can be 
used to identify the individual bands whenever only a single core 
configuration is taken into account. 

\subsection{The Hamiltonian}

The phenomenological SACM Hamiltonian is expressed in terms of boson 
number conserving 
combinations of the group generators. In the simplest applications 
it is sufficient to consider terms up to second order \cite{sacm}. A 
further assumption is considering only terms constructed from the 
Casimir invariants of the corresponding groups. This is called 
the dynamical symmetry approximation. In the case of SU(3) dynamical 
symmetry one typically considers a Hamiltonian of the type  
\begin{eqnarray}
H& = &
\hbar \omega C_1({\rm U}(3))+\chi_R C_2({\rm SU}_R(3)) 
+\chi C_2({\rm SU}(3))  \nonumber\\
&&
+\theta K^2 +\beta C_2({\rm SO}(3))+E_0\ .
\label{hami}
\end{eqnarray}
Here the first term is an harmonic oscillator with the parameter 
$\hbar\omega$ determined in MeV by the mass number as 
$\hbar\omega=45 A^{-1/3}-25A^{-2/3}$ for light nuclei, 
while $E_0$ is a constant that sets the ground-state energy to 
zero. The 
$C_2({\rm SU}(3))$ Casimir operators correspond 
to the combinations of the respective quadrupole-quadrupole 
and squared angular momentum terms. 
$C_2({\rm SO}(3))$ is the Casimir operator of the SO(3) group and it 
is equivalent with the square of the angular momentum operator. 
The $K^2$ operator is used to generate $K$-band splitting in the 
Elliott model, i.e. to lift the degeneracy of states with the same $L$ 
and different $\kappa$ values \cite{naqvi}. It can be defined as the 
square of the $L_3$ component of the angular momentum, which is the 
projection of $L$ on the body-fixed symmetry axis of the nucleus. 
It can be written as a special 
combination of three rotational scalar interactions constructed from 
the $L$ angular and $Q$ quadrupole moment operators \cite{naqvi}. 
It was also shown that it is close to being diagonal in the Elliott basis, 
so it is customary to neglect its 
off-diagonal terms. This term splits the energy of states with the same 
SU(3) labels and angular momenta. 
Taking the SU(3) basis (\ref{basis}) 
with the asumptions mentioned above the Hamiltonian (\ref{hami}) 
is diagonal, and the eigenvalues 
are given by 
\begin{eqnarray}
E& = &
\hbar \omega n_{\pi }+\chi_R n_\pi(n_{\pi}+3) 
+\chi (\lambda^2+\mu^2+\lambda\mu+3\lambda+3\mu)  
\nonumber\\
&&
+\theta \kappa^2+\beta L(L+1) +E_0\ .
\label{ener}
\end{eqnarray}
It is notable that the eigenvalues of $C_2({\rm SU}(3))$ are largest 
for the leading SU(3) representation that corresponds to 
maximal deformation. Due to the nature of the nuclear forces such 
states are always located at low energy compared to other 
representations from the same shell (same $n_{\pi}$). Since 
these states always appear in the 
SACM basis, it means that clustering implies the presence of 
maximally deformed states in the SU(3) basis.  

The $\theta$ parameter of the $\kappa^2$ term is often considered 
to be dependent on $n_{\pi}$, or at least on $(-1)^{n_{\pi}}$ on 
grounds that bands with $K=2$ typically occur above $K=0$ bands 
for positive-parity levels, while for the negative-parity spectrum 
states with $J^{\pi}=3^-$ or $2^-$ appear lowest, indicating that 
bands with $K=3$ or 2 should lie below bands with $K=1$ and 0, 
respectively. Since such low-lying negative-parity bands typically 
appear in $sd$-shell nuclei, in previous applications of the SACM 
in this domain this situation was handled on the phenomenological 
level by considering $n_{\pi}$-, or parity-dependent $\kappa^2$ terms 
\cite{cunda,other}. 
It is also possible to take into account the 
different moments of inertia associated to different bands. For 
this one can consider $\beta=\beta_0/\langle C_2({\rm SU(3)})+3\rangle$, 
because the SU(3) Casimir operator also expresses the measure of the 
deformation. 

\subsection{Electromagnetic transitions}
\label{emtr}

Here we discuss only E2 and E1 transitions. The E2 transitions are 
generated by quadrupole operators assigned both to the relative motion 
and the internal structure of the clusters (in this case this means the 
core), which are rank-2 tensors in terms of the SO(3) rotational group. 
These operators are also SU(3) tensors with SU(3) 
character (1,1) and angular momentum 2. The electric 
quadrupole transitions containing one-body terms are thus generated by 
the two-parameter operator 
\begin{equation}
T^{({\rm E}2)}=q_RQ^{(1,1)2}_R + q_CQ^{(1,1)2}_C\ . 
\label{te2}
\end{equation}
The selection rules of this operator are as follows. It allows only 
intrashell transitions, i.e. $n_{\pi}$ is conserved. Besides in-band 
transitions it allows changing $\kappa$ by 2 units and leaving 
$\lambda$ and $\mu$ unchanged. 
Transitions with $\Delta \lambda=\pm 2$ 
and $\Delta \mu=\mp 1$ are also allowed, although they are typically 
much weaker than in-band transitions. 
When $q_R=q_C$ the operator corresponds 
to the quadrupole momentum of the joint SU(3) group. In this case 
(\ref{te2}) generates transitions only within the same SU(3) 
representations, so only $\kappa$ can change. 

In order to describe transitions 
between major shells with $\Delta n_{\pi}=\pm 2$ it is necessary to 
add two-body terms to (\ref{te2}) acting only on the relative motion 
part of the states. 
(The excitations of the core nucleus are restricted 
to states belonging to the ground-state SU$_C$(3) representation, 
so excited-shell core states are missing fom the model space.) 
This operator can be constructed as 
\begin{equation}
T^{({\rm E}2)}_{\rm intershell}
 =p_R \left\{ 
  \left[[\sigma^{\dagger}\times\tilde{\pi}]^{(0,1)}
  \times
  [\sigma^{\dagger}\times\tilde{\pi}]^{(0,1)}\right]^{(0,2)2}
 + 
  \left[[\tilde{\sigma}\times\pi^{\dagger}]^{(1,0)}
  \times
  [\tilde{\sigma}\times\pi^{\dagger}]^{(1,0)}\right]^{(2,0)2}
 \right\}
\label{TE2_2hw}
\end{equation}
and its $SU_R(3)$ character  
is $(\lambda_R,\mu_R)=(0,2)$ for transitions with 
$\Delta n_{\pi}=-2$ and $(\lambda_R,\mu_R)=(2,0)$ for transitions with 
$\Delta n_{\pi}=2$. The selection rules for the two processes are 
given by the appropriate SU(3) multiplication rule \cite{sacm}, i.e. strong 
transitions are expected for $\Delta n_{\pi}=\pm 2$, $\Delta \lambda=\pm 2$, 
$\Delta\mu=0$, while weaker transitions with 
$\Delta n_{\pi}=\pm 2$, $\Delta \lambda=0$, $\Delta\mu=\pm 1$. 

The electric dipole operator connects shells with $\Delta n_{\pi}=\pm 1$. 
It is not an SU(3) tensor, rather it has mixed SU(3) character 
of $(\lambda,\mu)=(1,0)$ and (0,1):  
\begin{equation}
T^{({\rm E}1)}
 =d_R  
  \left[
  [\sigma^{\dagger}\times\tilde{\pi}]^{(0,1)}
 + 
  [\tilde{\sigma}\times\pi^{\dagger}]^{(1,0)}
  \right]\ .
\label{TE1}
\end{equation}
For $\Delta n_{\pi}=\mp 1$ it 
generates strong transitions with $\Delta\lambda=\mp 1$ and $\Delta\mu =0$ 
and weaker transitions with $\Delta\lambda=\pm 1$ and $\Delta\mu =\mp 1$ 
\cite{sacm}. 

The transition matrix elements composed of the transition operators 
(e.g. (\ref{te2})) and the basis states (\ref{basis}) can be calculated 
using SU(3) and 
SO(3) tensor algebraic manipulations that include SO(3) $6j$ 
coefficients, SU(3) $9j$ coefficients, SU(3) isoscalar factors and reduced 
matrix elements. See e.g. Ref. \cite{jpg07} for the details.

\section{Application to $^{22}$Ne}
\label{22ne_appl}

In our approach we assume that the Hamiltonian has
dynamical symmetry, so the energy eigenvalues can be calculated 
exactly. This also means that the spectrum contains rotational 
bands with energy dependence of the type $E\sim J(J+1)$. 
Obviously, this is not the case in general: typically only 
the most characteristic cluster bands (with $K=0$) show clear rotational 
pattern in the experimental energy spectrum. In a more realistic 
calculation, when the Hamiltonian is not diagonal one would expect 
such a result. The application of the SACM with broken SU(3) dynamical 
symmetry can be found in Ref. \cite{brokensu3}. 
The deviation from the rotational pattern would be 
more pronounced for bands that contain unnatural-parity states, 
because the composition of these states from various SU(3) 
representations would be different from that of natural-parity 
states. This is because unnatural-parity states appear only in 
basis states with $\mu\ne 0$, while natural-parity states appear 
in any $(\lambda,\mu)$ representation. 

The core nucleus for $^{22}$Ne is $^{18}$O. In the Elliott scheme its 
ground-state configuration is described by the (4,0) SU$_C$(3) 
representation that contains the $0^+$ ground state and the first 
excited $2^+(1.98)$ and $4^+(3.55)$ states. 
The spin-isospin structure is $T=1$ 
and $S=0$, which means that the orbital angular momentum is 
responsible for the $J$ angular momentum of the nucleus. 
The model space for $^{22}$Ne is displayed in Table \ref{ne22su3}. 

\begin{table}
\caption{SU(3) representations contained in the SACM model space 
of $^{22}{\rm Ne}\sim\alpha+^{18}{\rm O}$ up to $4\hbar\omega$ excitations 
(i.e. up to $n_{\pi}=12$). }
\begin{tabular}{rrrrrrr}
&&&&&\\
 $n_{\pi}$ & & & $(\lambda,\mu)$ & & \\
&&&&&\\
\hline
&&&&&\\
 8 & & & (8,2) & (6,3) & (4,4) \\
 9 & & (11,1) & (9,2) & (7,3) & (5,4) \\
 10 & (14,0) & (12,1) & (10,2) & (8,3) & (6,4)\\
 11 & (15,0) & (13,1) & (11,2) & (9,3) & (7,4)\\
 12 & (16,0) & (14,1) & (12,2) & (10,3) & (8,4)\\
&&&&&\\
\hline
\end{tabular}
\label{ne22su3}
\end{table}
\vskip .4cm

\subsection{The band structure}

The ground-state region should be dominated by configurations with 
$n_{\pi}=8$, i.e. $0\hbar\omega$ excitations. Of these the dominant one 
is expected to lie lowest in energy, which is the (8,2) representation. 
This contains two bands with $\kappa=0$ and 2, of which the latter one 
is expected to lie higher, similarly to other neighboring nuclei (e.g. 
$^{24}$Mg). The next positive-parity band in energy could be from the 
(6,3) SU(3) representation, which may have a lower-lying $\kappa=1$ and 
a higher-lying $\kappa=3$ band. 
We note here that in the Elliott model the model space of $^{22}$Ne 
with the appropriate permutational structure 
contains 13 SU(3) representations for $0\hbar\omega$ i.e. 
$n_{\pi}=8$, some with multiplicities \cite{harvey}. The SACM model 
space contains only three of them (see Table \ref{ne22su3}), including 
the two leading ones, (8,2) and (6,3) with the largest deformation,  
which are expected to be located lowest in excitation energy. 

The low-lying negative-parity spectrum should contain states from the 
leading (11,1) representation, which correspods to nearly prolate 
deformation and has a single band with $\kappa=1$. States from the 
(9,2) with $\kappa=0$ and 2 should also lie close. The deformation 
of these states is similar to that of the (8,2) configuration in the 
ground-state region. The difference is that in contrast with the 
positive-parity spectrum, here the $\kappa=2$ band is expected to be 
below the $\kappa=0$ band. 

Somewhat higher the $2\hbar\omega$ states should appear with 
$n_{\pi}=10$. The leading SU(3) representation here is (14,0), which 
corresponds to a highly deformed prolate structure with states 
forming a $\kappa=0$ rotational band. Further up in energy the states 
from the (12,1) representation with $\kappa=1$ are also expected to 
appear. This configuration is also close to a highly deformed 
prolate shape. 

The negative-parity correspondent of the (14,0) band is (15,0), 
which is also a highly deformed configuration composed of states 
belonging to a $\kappa=0$ band. 

Further positive-parity $\kappa=0$ bands may appear in or above this 
region with $n_{\pi}=8$ $(\lambda,\mu)=(4,4)$ or 
$n_{\pi}=10$ $(\lambda,\mu)=(10,2)$, depending on the parameters of $H$. 
The former one would be a compact, triaxial structure, while the 
latter one would correspond to a close to prolate shape with 
relatively large deformation. 

In assigning the experimental states to bands one can use the 
relative position of the energy levels (e.g. rotation structures) 
and electromagnetic transition data. We combine these with the 
band assignment of other works \cite{kimura,desc88}, making an 
effort to assign all the low-lying experimental states 
to model ones for every $J^{\pi}$. 
The band structure we apply in the 
case of $^{22}$Ne is displayed in Tables \ref{bandassp} and \ref{bandassn} 
for positive- and negative-parity states, respectively. 
We used experimental 
data taken from the Brookhaven National Laboratory data base \cite{bnl}. 

\begin{table}[h]
\caption{Assignment of positive-parity experimental $^{22}$Ne states to 
the SACM 
states. $E_{\rm Th.}$ is the result of a simple fit described in 
Subsection \ref{22nespectr}. States with $J\le 8$ are displayed. 
For bands expected to lie higher only the lowest member of the 
band is indicated.}
\begin{tabular}{rrrrr}
$n_{\pi}(\lambda,\mu)\kappa$ & $J^{\pi}$ & $E_{\rm Th.}$ & $K^{\pi}$ 
 & $J^{\pi}(E)_{\rm Exp.}$ \\
&& (MeV) && (MeV) \\
\hline
8(8,2)0 & $0^+$ & 0 & $0^+_1$ & $0^+(0)$ \\
 & $2^+$ & 0.865 & & $2^+(1.275)$ \\
 & $4^+$ & 2.882 & & $4^+(3.358)$ \\
 & $6^+$ & 6.053 & & $(6^+)(6.310)$ \\
 & $8^+$ & 10.377 & & $(8^+)(11.032)$ \\
8(8,2)2 & $2^+$ & 4.349 & $2^+$ & $2^+(4.456)$ \\
 & $3^+$ & 5.214 & & $3^+(5.641)$ \\
 & $4^+$ & 6.367 & & $(4)^+(5.524)$ \\
 & $5^+$ & 7.808 & & $(5^+)(7.422)$ \\
 & $6^+$ & 9.538 & & \\
8(6,3)1 & $1^+$ & 4.404 & $1^+$ & $1^+(5.332)$ \\
 & $2^+$ & 5.129 & & $2^+(5.363)$ \\
 & $3^+$ & 6.217 & & $3^+(6.635)$ \\
 & $4^+$ & 7.668 & & $4^+(6.347)$ \\
 & $5^+$ & 9.481 & & \\
8(6,3)3 & $3^+$ & 14.985 & & \\
8(4,4)0 & $0^+$ & 5.154 & & $0^+(7.341)$ \\
 & $2^+$ & 6.503 & & \\
 & $4^+$ & 9.651 & & \\
8(4,4)2 & $2^+$ & 11.940 & & \\
8(4,4)4 & $4^+$ & 31.936 & & \\
10(14,0)0 & $0^+$ & 6.230 & $0^+_2$ & $0^+(6.234)$ \\
 & $2^+$ & 6.650 & & $2^+(7.665)$ \\
 & $4^+$ & 7.630 & & $(4)^+(8.076)$ \\
 & $6^+$ & 9.169 & & \\
 & $8^+$ & 11.268 & & \\
10(12,1)1 & $1^+$ & 12.067 & & \\
10(10,2)0 & $0^+$ & 15.803 & & \\
12(16,0)0 & $0^+$ & 18.758 & & \\
\hline
\end{tabular}
\label{bandassp}
\end{table}
\vskip .4cm

\begin{table}[h]
\caption{The same as Table \ref{bandassp} for negative-parity states of 
$^{22}$Ne. States with $J\le 7$ are displayed. }
\begin{tabular}{rrrrr}
$n_{\pi}(\lambda,\mu)\kappa$ & $J^{\pi}$ & $E_{\rm Th.}$ & $K^{\pi}$ 
 & $J^{\pi}(E)_{\rm Exp.}$ \\
&& (MeV) && (MeV) \\
\hline
9(11,1)1 & $1^-$ & 3.938 & $1^-$ & $1^-(7.051)$ \\
 & $2^-$ & 4.330 & & $2^-(7.665)$ \\
 & $3^-$ & 4.918 & & $(3)^-(8.376)$ \\
 & $4^-$ & 5.702 & & \\
 & $5^-$ & 6.682 & & \\
9(9,2)0 & $1^-$ & 8.370 & $0^-_1$ & $1^-(6.691)$ \\
 & $3^-$ & 9.583 & & $3^-(7.406)$ \\
 & $5^-$ & 11.766 & & \\
 & $7^-$ & 14.920 & & \\
9(9,2)2 & $2^-$ & 7.194 & $2^-$ & $2^-(5.146)$ \\
 & $3^-$ & 7.922 & & $3^-(5.910)$ \\
 & $4^-$ & 8.892 & & \\
 & $5^-$ & 10.105 & & \\
9(7,3)1 & $1^-$ & 11.226 & & \\
9(7,3)3 & $3^-$ & 8.609 & & $3^-(7.722)$ \\
9(5,4)0 & $1^-$ & 14.388 & & \\
9(5,4)2 & $2^-$ & 12.593 & & \\
9(5,4)4 & $4^-$ & 7.567 & & $(3)^-(8.376)$ \\
11(15,0)0 & $1^-$ & 12.843 & $0^-_2$ & $1^-(12)$ \\
 & $3^-$ & 13.461 & & \\
 & $5^-$ & 14.572 & & \\
 & $7^-$ & 16.178 & & \\
11(13,1)1 & $1^-$ & 18.137 & & \\
\hline
\end{tabular}
\label{bandassn}
\end{table}
\vskip .4cm

\subsection{The energy spectrum}
\label{22nespectr}

\begin{figure}

\unitlength=1.cm
\begin{picture}(10,15)(0,0)

\put(0,0){\line(1,0){10}}
\put(0,15){\line(1,0){10}}
\put(0,0){\line(0,1){15}}
\put(10,0){\line(0,1){15}}

\put(0,2.5){\line(1,0){0.2}}
\put(-0.3,2.5){\makebox(0,0){10}}
\put(10,2.5){\line(-1,0){0.2}}

\put(0,5){\line(1,0){0.2}}
\put(-0.3,5){\makebox(0,0){20}}
\put(10,5){\line(-1,0){0.2}}

\put(0,7.5){\line(1,0){0.2}}
\put(-0.3,7.5){\makebox(0,0){30}}
\put(10,7.5){\line(-1,0){0.2}}

\put(0,10){\line(1,0){0.2}}
\put(-0.3,10){\makebox(0,0){40}}
\put(10,10){\line(-1,0){0.2}}

\put(0,12.5){\line(1,0){0.2}}
\put(-0.3,12.5){\makebox(0,0){50}}
\put(10,12.5){\line(-1,0){0.2}}

\put(-0.5,14){\makebox(0,0){$E_x$}}
\put(-0.6,13.5){\makebox(0,0){(MeV)}}



\put(0,-0.3){\makebox(0,0){0}}
\put(10,-0.3){\makebox(0,0){0.2}}
\put(5,-0.5){\makebox(0,0){$-\chi$}}

\put(10.7,-0.1){\makebox(0,0){8(8,2)}}

\qbezier(0,0)(5,0.6)(10,1.2)
\put(10.7,1.2){\makebox(0,0){8(6,3)}}

\qbezier(0,0)(5,1.05)(10,2.1)
\put(10.7,2.1){\makebox(0,0){8(4,4)}}

\qbezier[100](0,3.22)(5,1.845)(10,0.47)
\put(12,0.47){\makebox(0,0){9(11,1)}}

\qbezier[100](0,3.22)(5,2.67)(10,2.12)
\put(12,2.12){\makebox(0,0){9(9,2)}}

\qbezier[100](0,3.22)(5,3.345)(10,3.47)
\put(12,3.47){\makebox(0,0){9(7,3)}}

\qbezier[100](0,3.22)(5,3.87)(10,4.52)
\put(12,4.52){\makebox(0,0){9(5,4)}}

\qbezier(0,6.44)(5,3.34)(10,0.24)
\put(10.7,0.24){\makebox(0,0){10(14,0)}}

\qbezier(0,6.44)(5,4.39)(10,2.34)
\put(10.7,2.44){\makebox(0,0){10(12,1)}}

\qbezier(0,6.44)(5,5.29)(10,4.14)
\put(10.7,4.14){\makebox(0,0){10(10,2)}}

\qbezier(0,6.44)(5,6.04)(10,5.64)
\put(10.7,5.44){\makebox(0,0){10(8,3)}}

\qbezier(0,6.44)(5,6.64)(10,6.84)
\put(10.7,6.84){\makebox(0,0){10(6,4)}}

\qbezier[100](0,9.66)(5,5.76)(10,1.86)
\put(12,1.76){\makebox(0,0){11(15,0)}}

\qbezier[100](0,9.66)(5,6.885)(10,4.11)
\put(12,4.11){\makebox(0,0){11(13,1)}}

\qbezier[100](0,9.66)(5,7.86)(10,6.06)
\put(12,6.06){\makebox(0,0){11(11,2)}}

\qbezier[100](0,9.66)(5,8.685)(10,7.71)
\put(12,7.71){\makebox(0,0){11(9,3)}}

\qbezier[100](0,9.66)(5,9.36)(10,9.06)
\put(12,9.06){\makebox(0,0){11(7,4)}}

\qbezier(0,12.88)(5,8.13)(10,3.38)
\put(10.7,3.38){\makebox(0,0){12(16,0)}}

\qbezier(0,12.88)(5,9.33)(10,5.78)
\put(10.7,5.78){\makebox(0,0){12(14,1)}}

\qbezier(0,12.88)(5,10.38)(10,7.88)
\put(10.7,7.88){\makebox(0,0){12(12,2)}}

\qbezier(0,12.88)(5,11.28)(10,9.68)
\put(10.7,9.68){\makebox(0,0){12(10,3)}}

\qbezier(0,12.88)(5,12.03)(10,11.18)
\put(10.7,11.18){\makebox(0,0){12(8,4)}}

\end{picture}

\caption{Estimation of the band-head energies as the function of the 
$\chi$ parameter with even (full line) and odd (dotted line) parity. 
To the right the quantum numbers $n_{\pi}(\lambda,\mu)$  
are displayed in separate columns for even- and odd-parity bands. 
Bands with $n_{\pi}\le 12$ are displayed. 
The lowest band with $n_{\pi}=13$ would appear at $E_x=19.2$ MeV 
for $\chi=-0.2$.}

\label{ne22bh}
\end{figure}
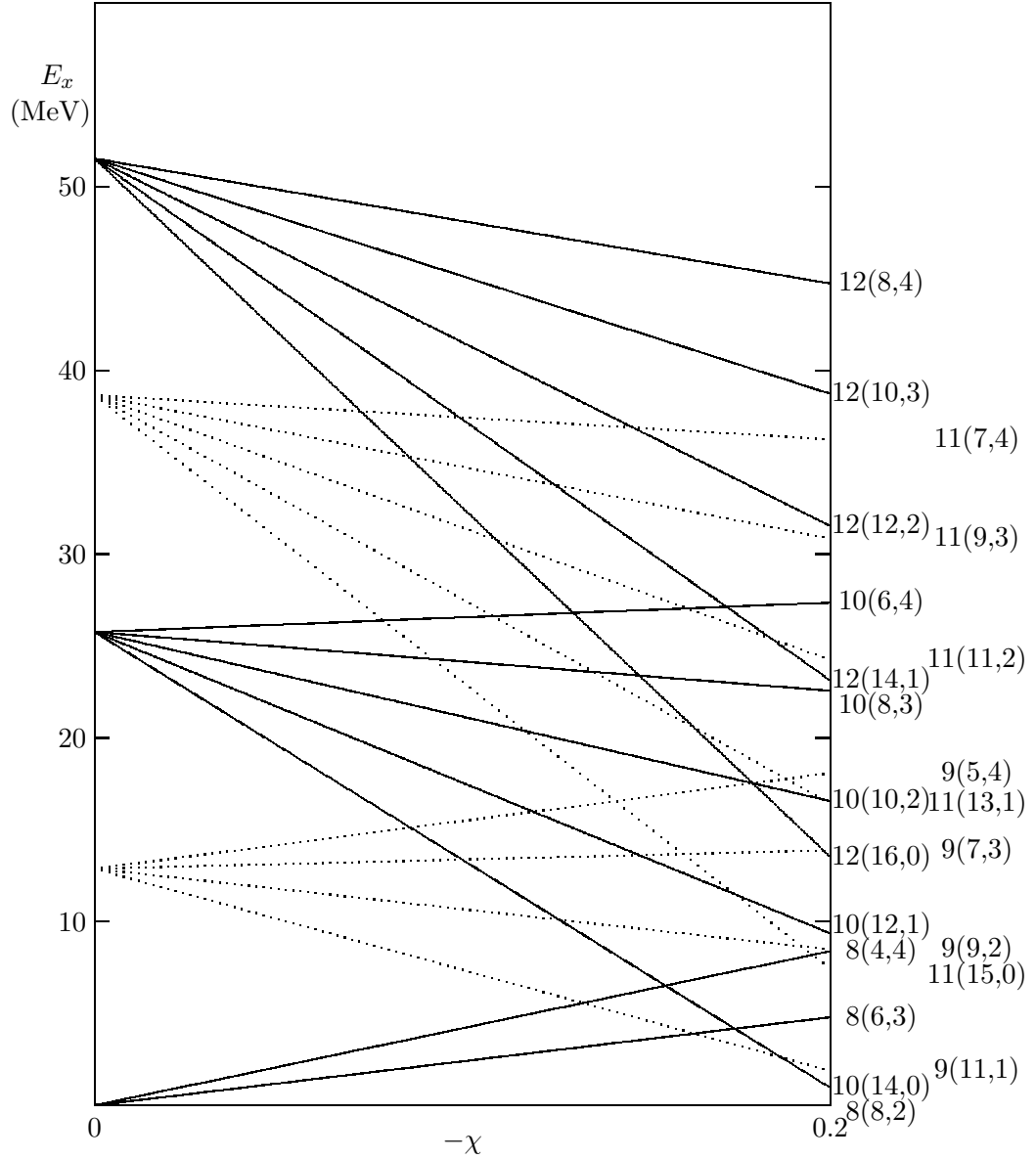

We apply (\ref{ener}) to fit the energy eigenvalues. The oscillator 
constant for $A=22$  is $\hbar\omega=12.88$ MeV according to the 
formula cited after Eq. (\ref{hami}). If the anharmonic $C_2({\rm SU}_R(3))$ 
term associated to the relative motion is neglected by setting $\chi_R=0$,  
the band-head energies can be estimated 
from the graph displayed in Figure \ref{ne22bh}, where the 
location of various SU(3) states is plotted as the function of 
parameter $\chi$. The actual band-head energies are also influenced 
by the $\theta \kappa^2$ term for bands with $\kappa\ne 0$, and 
to a small extent also by the rotational term $\beta L(L+1)$ when 
$L\ne 0$ holds for the band-head state, nevertheless, Fig. 
\ref{ne22bh} gives 
reasonable support to estimate the sequence of bands. 

%
%

The assignment of experimental levels to model states is contained 
in Tables \ref{bandassp} and \ref{bandassn} for positive- and 
negative-parity states, respectively. 

The $\beta$ parameter sets the steepness of the rotational bands. 
Since this depends on the moment of inertia, which, in turn depends 
on the deformation, we parametrized it as 
$\beta=\beta_0/(\lambda^2+\mu^2+\lambda\mu+3\lambda+3\mu+3)$, 
taking into account the eigenvalues of the  $C_2({\rm SU(3))}$ operator 
for each band. 

The $\kappa^2$ term seems to exhibit parity 
dependence, as the $2^+$ band appears above the ground-state 
$0^+_1$ band, while for the negative-parity levels the $2^-$ 
band is the lowest. For this reason we consider different values 
of $\theta$ for positive- ($\theta_+$) and negative-parity 
($\theta_-$) bands. Similarly to the rotational constant $\beta$ we 
also parametrized the $\kappa$-dependent terms with the factor 
$(\lambda^2+\mu^2+\lambda\mu+3\lambda+3\mu+3)^{-1}$ reflecting the 
effect of deformation. 

In conclusion, there are four parameters that should be determined 
from the fit to the experimental data: $\chi$, $\theta_+$, $\theta_-$ 
and $\beta_0$. Of these the first three determine the location of the 
band heads (although $\beta_0$ also has  a slight influence on 
bands with $\kappa\ne 0$). Since in the fit we wished to consider 
also bands belonging to $3\hbar\omega$ and $4\hbar\omega$ 
excitations ($n_{\pi}=3$ and 4), we also included the anharmonic 
term associated with the relative motion. Allowing $\chi_R\ne 0$ 
in Eq. (\ref{ener}) modifies the relative position of bands 
with different $n_{\pi}$, but has no direct influence on the 
relative position of bands (SU(3) multiplets) with the same 
$n_{\pi}$. 

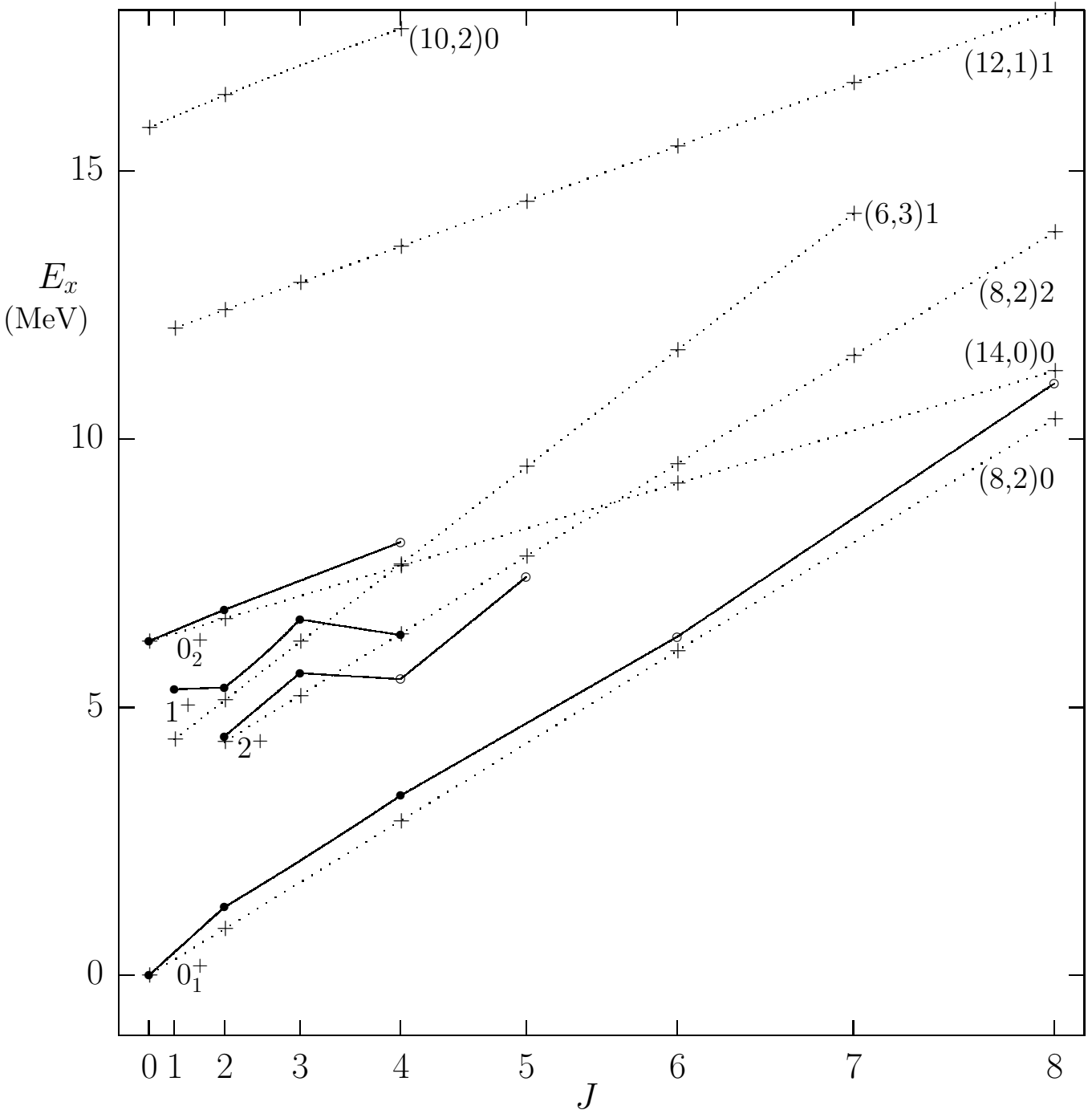
\begin{figure}[h]
\setlength{\unitlength}{1cm}
\begin{picture}(  16.00,  17.00)(0,-2)
\put(   0.00,-1){\line(1,0){  16.00}}
\put(   0.00,  16.00){\line(1,0){  16.00}}
\put(   0.00,-1){\line(0,1){  17.00}}
\put(  16.00,-1){\line(0,1){  17.00}}
\put(   0.00,   0.00){\line(1,0){   0.25}}
\put(  16.00,   0.00){\line(-1,0){   0.25}}
\put(  -0.25,   0.05){\makebox(0,0)[r]{\Large{$  0$}}}
\put(   0.00,   4.44){\line(1,0){   0.25}}
\put(  16.00,   4.44){\line(-1,0){   0.25}}
\put(  -0.25,   4.49){\makebox(0,0)[r]{\Large{$  5$}}}
\put(   0.00,   8.89){\line(1,0){   0.25}}
\put(  16.00,   8.89){\line(-1,0){   0.25}}
\put(  -0.25,   8.94){\makebox(0,0)[r]{\Large{$ 10$}}}
\put(   0.00,  13.33){\line(1,0){   0.25}}
\put(  16.00,  13.33){\line(-1,0){   0.25}}
\put(  -0.25,  13.38){\makebox(0,0)[r]{\Large{$ 15$}}}
\put(   0.50,-1){\line(0,1){0.25}}
\put(   0.50,  16.00){\line(0,-1){0.25}}
\put(   0.65,  -1.50){\makebox(0,0)[r]{\Large{$  0$}}}
\put(   0.92,-1){\line(0,1){0.25}}
\put(   0.92,  16.00){\line(0,-1){0.25}}
\put(   1.07,  -1.50){\makebox(0,0)[r]{\Large{$  1$}}}
\put(   1.75,-1){\line(0,1){0.25}}
\put(   1.75,  16.00){\line(0,-1){0.25}}
\put(   1.90,  -1.50){\makebox(0,0)[r]{\Large{$  2$}}}
\put(   3.00,-1){\line(0,1){0.25}}
\put(   3.00,  16.00){\line(0,-1){0.25}}
\put(   3.15,  -1.50){\makebox(0,0)[r]{\Large{$  3$}}}
\put(   4.67,-1){\line(0,1){0.25}}
\put(   4.67,  16.00){\line(0,-1){0.25}}
\put(   4.82,  -1.50){\makebox(0,0)[r]{\Large{$  4$}}}
\put(   6.75,-1){\line(0,1){0.25}}
\put(   6.75,  16.00){\line(0,-1){0.25}}
\put(   6.90,  -1.50){\makebox(0,0)[r]{\Large{$  5$}}}
\put(   9.25,-1){\line(0,1){0.25}}
\put(   9.25,  16.00){\line(0,-1){0.25}}
\put(   9.40,  -1.50){\makebox(0,0)[r]{\Large{$  6$}}}
\put(  12.17,-1){\line(0,1){0.25}}
\put(  12.17,  16.00){\line(0,-1){0.25}}
\put(  12.32,  -1.50){\makebox(0,0)[r]{\Large{$  7$}}}
\put(  15.50,-1){\line(0,1){0.25}}
\put(  15.50,  16.00){\line(0,-1){0.25}}
\put(  15.65,  -1.50){\makebox(0,0)[r]{\Large{$  8$}}}
\put(   7.75,-2.0){\makebox(0,0){\Large{$J$}}}
\put(  -1.00,  11.56){\makebox(0,0){\Large{$E_x$}}}
\put(  -1.20,  10.86){\makebox(0,0){\large{(MeV)}}}
\put(   0.37,  -0.09){$+$}
\put(   1.62,   0.68){$+$}
\put(   4.54,   2.47){$+$}
\put(   9.12,   5.29){$+$}
\put(  15.37,   9.13){$+$}
\qbezier[100](0.5,0.0)(8.0,4.61)(15.5,9.22)
\put(   15.5,8.2){\makebox(0,0)[r]{\large{(8,2)0}}}
\put(   0.50,   0.00){\circle*{.15}}
\put(   1.75,   1.13){\circle*{.15}}
\qbezier( 0.50,   0.00)(1.125, 0.565)( 1.75,   1.13 )
\put(   4.67,   2.98){\circle*{.15}}
\qbezier( 1.75,   1.13 )(3.21, 2.005)(  4.67,   2.98 )
\put(   9.25,   5.61){\circle{.15}}
\qbezier( 4.67,   2.98 )(6.96, 4.295)(  9.25,   5.61  )
\put(  15.50,   9.81){\circle{.15}}
\qbezier(  9.25,   5.61 )(12.375, 7.71)( 15.50,   9.81  )
\put(  1.5,0.0){\makebox(0,0)[r]{\large{$0^+_1$}}}

\put(   1.62,   3.78){$+$}
\put(   2.87,   4.54){$+$}
\put(   4.54,   5.57){$+$}
\put(   6.62,   6.85){$+$}
\put(   9.12,   8.39){$+$}
\put(  12.04,  10.18){$+$}
\put(  15.37,  12.23){$+$}
\qbezier[100](  1.75,   3.87)(8.625,8.095)(  15.5,  12.32)
\put(   15.5,11.3){\makebox(0,0)[r]{\large{(8,2)2}}}
\put(   1.75,   3.96){\circle*{.15}}
\put(   3.00,   5.01){\circle*{.15}}
\qbezier( 1.75,3.96)(2.375,4.485)( 3.0,5.01 )
\put(   4.67,   4.91){\circle{.15}}
\qbezier(3.0,5.01 )(3.835, 4.96)(4.67,   4.91 )
\put(   6.75,   6.60){\circle{.15}}
\qbezier(4.67, 4.91)(5.71, 5.755 )(6.75, 6.6)
\put(  2.5,3.8){\makebox(0,0)[r]{\large{$2^+$}}}

\put(   0.79,   3.82){$+$}
\put(   1.62,   4.47){$+$}
\put(   2.87,   5.44){$+$}
\put(   4.54,   6.73){$+$}
\put(   6.62,   8.34){$+$}
\put(   9.12,  10.27){$+$}
\put(  12.04,  12.53){$+$}
\qbezier[100](   0.92,   3.91)(6.545,8.265)( 12.17, 12.62)
\put(   13.6,12.6){\makebox(0,0)[r]{\large{(6,3)1}}}
\put(   0.92,   4.74){\circle*{.15}}
\put(   1.75,   4.77){\circle*{.15}}
\qbezier( 0.92, 4.74 )(1.335, 4.755 )( 1.75, 4.77)
\put(   3.00,   5.90){\circle*{.15}}
\qbezier( 1.75, 4.77 )(2.475, 5.335)(3.0, 5.9)
\put(   4.67,   5.64){\circle*{.15}}
\qbezier( 3.0, 5.9 )(3.835, 5.77)(4.67, 5.64)
\put(  1.3,4.4){\makebox(0,0)[r]{\large{$1^+$}}}




\put(   0.37,   5.45){$+$}
\put(   1.62,   5.82){$+$}
\put(   4.54,   6.69){$+$}
\put(   9.12,   8.06){$+$}
\put(  15.37,   9.93){$+$}
\qbezier[100](   0.5,   5.54 )(8.0,7.78)(15.5,   10.02 )
\put(   15.5,10.3){\makebox(0,0)[r]{\large{(14,0)0}}}
\put(   0.50,   5.54){\circle*{.15}}
\put(   1.75,   6.06){\circle*{.15}}
\qbezier(0.5, 5.54 )(1.125, 5.8)(1.75, 6.06)
\put(   4.67,   7.18){\circle{.15}}
\qbezier(1.75, 6.06)(3.21, 6.62)(4.67, 7.18)
\put(  1.5,5.4){\makebox(0,0)[r]{\large{$0^+_2$}}}

\put(   0.79,  10.64){$+$}
\put(   1.62,  10.94){$+$}
\put(   2.87,  11.39){$+$}
\put(   4.54,  11.99){$+$}
\put(   6.62,  12.74){$+$}
\put(   9.12,  13.65){$+$}
\put(  12.04,  14.70){$+$}
\put(  15.37,  15.91){$+$}
\qbezier[100](  0.92,  10.73 )(8.21,13.365)(  15.5,  16.0)
\put(   15.5,15.1){\makebox(0,0)[r]{\large{(12,1)1}}}

\put(   0.37,  13.96){$+$}
\put(   1.62,  14.51){$+$}
\put(   4.54,  15.60){$+$}
\qbezier[40]( 0.5,  14.05  )(2.585,14.97)(   4.67,  15.69 )
\put(   6.3,15.5){\makebox(0,0)[r]{\large{(10,2)0}}}
\end{picture}

\caption{Positive-parity bands displayed in a rotational diagram form. 
Experimental bands are indicated with full line and $K^{\pi}$, 
while those predicted by the SACM are marked with dotted line and 
the $(\lambda,\mu)\kappa$ labels.Open circles denote experimental 
states with uncertain $J^{\pi}$ assignment.}
\label{rot-p}
\end{figure}
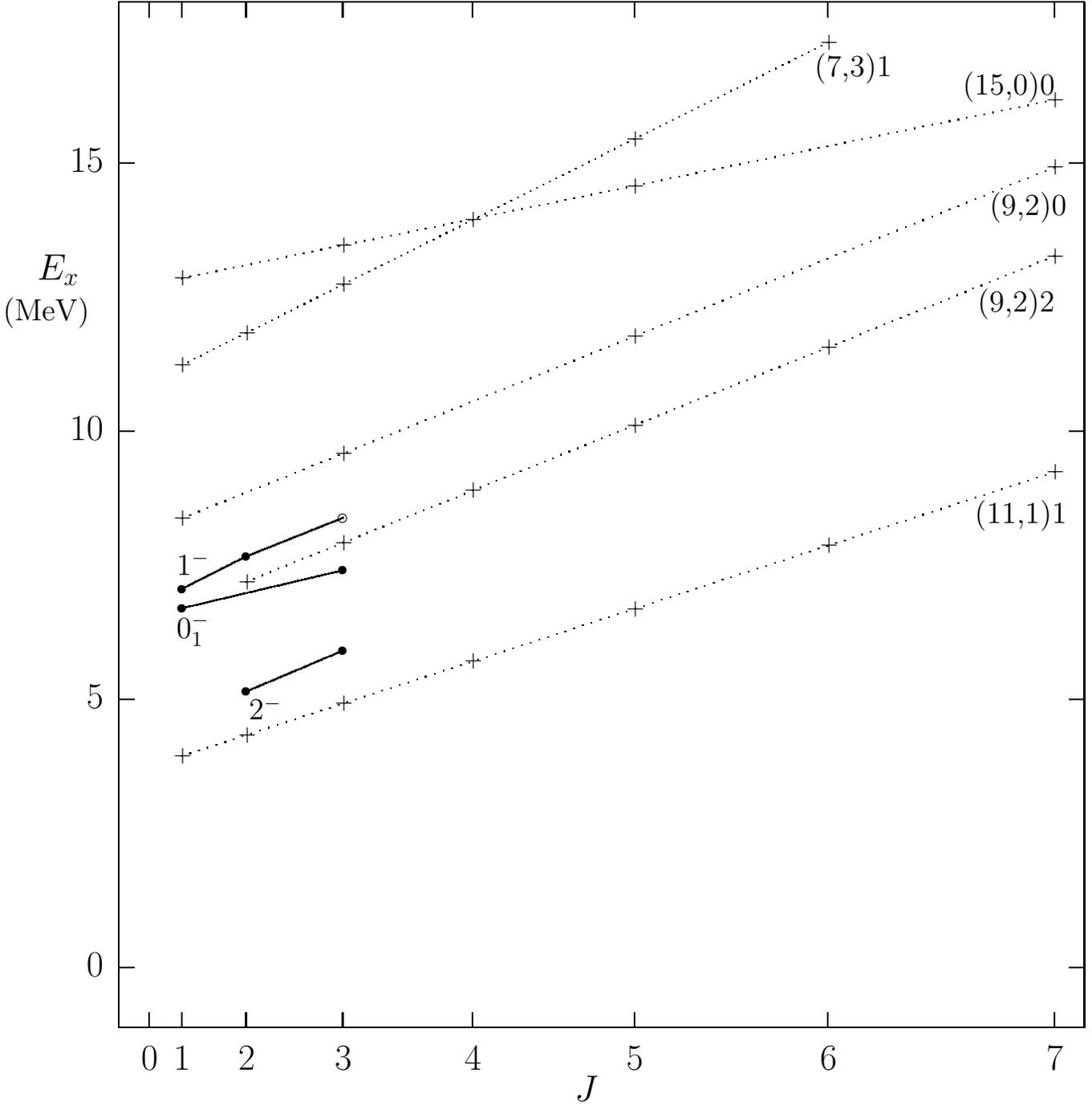
\begin{figure}[h]
\setlength{\unitlength}{1cm}
\begin{picture}(  16.00,  17.00)(0,-2)
\put(   0.00,-1){\line(1,0){  16.00}}
\put(   0.00,  16.00){\line(1,0){  16.00}}
\put(   0.00,-1){\line(0,1){  17.00}}
\put(  16.00,-1){\line(0,1){  17.00}}
\put(   0.00,   0.00){\line(1,0){   0.25}}
\put(  16.00,   0.00){\line(-1,0){   0.25}}
\put(  -0.25,   0.05){\makebox(0,0)[r]{\Large{$  0$}}}
\put(   0.00,   4.44){\line(1,0){   0.25}}
\put(  16.00,   4.44){\line(-1,0){   0.25}}
\put(  -0.25,   4.49){\makebox(0,0)[r]{\Large{$  5$}}}
\put(   0.00,   8.89){\line(1,0){   0.25}}
\put(  16.00,   8.89){\line(-1,0){   0.25}}
\put(  -0.25,   8.94){\makebox(0,0)[r]{\Large{$ 10$}}}
\put(   0.00,  13.33){\line(1,0){   0.25}}
\put(  16.00,  13.33){\line(-1,0){   0.25}}
\put(  -0.25,  13.38){\makebox(0,0)[r]{\Large{$ 15$}}}
\put(   0.50,-1){\line(0,1){0.25}}
\put(   0.50,  16.00){\line(0,-1){0.25}}
\put(   0.65,  -1.50){\makebox(0,0)[r]{\Large{$  0$}}}
\put(   1.04,-1){\line(0,1){0.25}}
\put(   1.04,  16.00){\line(0,-1){0.25}}
\put(   1.19,  -1.50){\makebox(0,0)[r]{\Large{$  1$}}}
\put(   2.11,-1){\line(0,1){0.25}}
\put(   2.11,  16.00){\line(0,-1){0.25}}
\put(   2.26,  -1.50){\makebox(0,0)[r]{\Large{$  2$}}}
\put(   3.71,-1){\line(0,1){0.25}}
\put(   3.71,  16.00){\line(0,-1){0.25}}
\put(   3.86,  -1.50){\makebox(0,0)[r]{\Large{$  3$}}}
\put(   5.86,-1){\line(0,1){0.25}}
\put(   5.86,  16.00){\line(0,-1){0.25}}
\put(   6.01,  -1.50){\makebox(0,0)[r]{\Large{$  4$}}}
\put(   8.54,-1){\line(0,1){0.25}}
\put(   8.54,  16.00){\line(0,-1){0.25}}
\put(   8.69,  -1.50){\makebox(0,0)[r]{\Large{$  5$}}}
\put(  11.75,-1){\line(0,1){0.25}}
\put(  11.75,  16.00){\line(0,-1){0.25}}
\put(  11.90,  -1.50){\makebox(0,0)[r]{\Large{$  6$}}}
\put(  15.50,-1){\line(0,1){0.25}}
\put(  15.50,  16.00){\line(0,-1){0.25}}
\put(  15.65,  -1.50){\makebox(0,0)[r]{\Large{$  7$}}}
\put(   7.75,-2.0){\makebox(0,0){\Large{$J$}}}
\put(  -1.00,  11.56){\makebox(0,0){\Large{$E_x$}}}
\put(  -1.20,  10.86){\makebox(0,0){\large{(MeV)}}}
\put(   0.91,   3.41){$+$}
\put(   1.98,   3.76){$+$}
\put(   3.58,   4.28){$+$}
\put(   5.73,   4.98){$+$}
\put(   8.41,   5.85){$+$}
\put(  11.62,   6.90){$+$}
\put(  15.37,   8.12){$+$}
\qbezier[100](1.04, 3.5)(8.27, 5.855)(15.5, 8.21)
\put(   15.7,7.5){\makebox(0,0)[r]{\large{(11,1)1}}}
\put(   1.04,   6.27){\circle*{.15}}
\put(   2.11,   6.81){\circle*{.15}}
\qbezier( 1.04, 6.27 )(1.575, 6.54)( 2.11, 6.81 )
\put(   3.71,   7.45){\circle{.15}}
\qbezier(  2.11, 6.81 )(2.91, 7.13)( 3.71, 7.45)
\put(  1.5,6.7){\makebox(0,0)[r]{\large{$1^-$}}}

\put(   0.91,   7.35){$+$}
\put(   3.58,   8.43){$+$}
\put(   8.41,  10.37){$+$}
\put(  15.37,  13.17){$+$}
\qbezier[100](1.04, 7.44)(8.27, 10.35)(15.5, 13.26)
\put(   15.7,12.6){\makebox(0,0)[r]{\large{(9,2)0}}}
\put(   1.04,   5.95){\circle*{.15}}
\put(   3.71,   6.58){\circle*{.15}}
\qbezier(  1.04, 5.95 )(2.375, 6.265)( 3.71, 6.58)
\put(  1.5,5.6){\makebox(0,0)[r]{\large{$0^-_1$}}}

\put(   1.98,   6.30){$+$}
\put(   3.58,   6.95){$+$}
\put(   5.73,   7.81){$+$}
\put(   8.41,   8.89){$+$}
\put(  11.62,  10.19){$+$}
\put(  15.37,  11.70){$+$}
\qbezier[100](2.11, 6.39)(8.805, 9.09)(15.5, 11.79)
\put(   15.5,11.0){\makebox(0,0)[r]{\large{(9,2)2}}}
\put(   2.11,   4.57){\circle*{.15}}
\put(   3.71,   5.25){\circle*{.15}}
\qbezier(  2.11, 4.57 )(2.91, 4.91)( 3.71, 5.25)
\put(  2.7,4.3){\makebox(0,0)[r]{\large{$2^-$}}}

\put(   0.91,   9.89){$+$}
\put(   1.98,  10.42){$+$}
\put(   3.58,  11.23){$+$}
\put(   5.73,  12.30){$+$}
\put(   8.41,  13.64){$+$}
\put(  11.62,  15.24){$+$}
\qbezier[100](1.04, 9.98)( 6.395, 12.655)(11.75, 15.33)
\put(   12.8,14.9){\makebox(0,0)[r]{\large{(7,3)1}}}





\put(   0.91,  11.33){$+$}
\put(   3.58,  11.88){$+$}
\put(   8.41,  12.86){$+$}
\put(  15.37,  14.29){$+$}
\qbezier[100](1.04, 11.42)(8.27, 12.9)(15.5, 14.38)
\put(   15.5,14.6){\makebox(0,0)[r]{\large{(15,0)0}}}

\end{picture}

\caption{The same as Fig. \ref{rot-p} for negative-parity bands.}
\label{rot-n}
\end{figure}

The fitted parameters (in MeV) are $\chi_R=-0.1027$, $\chi=-0.1227$, 
$\theta_+=101.929$, $\theta_-=-57.713$ and $\beta_0=18.862$.  
The calculated energy eigenvalues are displayed in Tables 
\ref{bandassp} and \ref{bandassn} along with the corresponding 
experimental energy values and the relevant $K^{\pi}$ 
band labels. These latter ones include labels generally 
accepted in the experimental assignment of levels, but also 
some assignments from theoretical works \cite{kimura,desc88}. 
The assignment of individual levels to bands in these works 
occasionally differs from our choice. The positive- and negative-parity 
energy spectrum is also displayed in a rotational diagram form in 
Figs. \ref{rot-p} and \ref{rot-n}, respectively. Note that the 
figures contain only model bands that correspond to well-established 
experimental bands, or can be candidates for bands predicted by 
other models \cite{kimura,desc88}. 

%
%

%
%

It is seen that the calculated positive-parity band structure reproduces 
the experimental levels rather well. The 8(8,2)2 and 8(6,3)1 bands 
have rotational structure in the calculations, while their experimental 
counterparts have even--odd staggering character. As we have mentioned 
earlier, such a feature could be reproduced in calculations with a 
Hamiltonian that does not have exact dynamical symmetry, and the energy 
eigenvalues would be calculated by diagonalization. Then the natural- 
and unnatural-parity states would have different composition that 
may lead to an even--odd staggering pattern. 

The band-head energy of the calculated negative-parity bands is 
somewhat different from those of the experimental ones. 
The calculations predict the 9(11,1)1 band lowest around $E_x=4$ MeV, 
about 3 MeV's lower than the experimental counterpart at 7.051 MeV. 
In contrast, the 9(9,2)2 and 9(9,2)0 bands lie about 2 MeV too high.  
It seems that the simple Hamiltonian (\ref{hami}) is not flexible enough 
to reproduce all the details of the band structure. The $\kappa$-dependent 
term in the negative-parity spectrum may also be too simple to get all the 
bands at the correct position. 

It is notable that bands corresponding to highly deformed prolate 
structure appear in the energy domain where molecular bands are 
expected to appear, i.e. around 12 to 18 MeV. These include the 
positive-parity (12,1)1 and (10,2)0 bands with $n_{\pi}=10$, the 
(16,0)0 band with $n_{\pi}=12$ and the negative-parity bands 
(15,0)0 and (13,1)1 with $n_{\pi}=11$ (see Tables \ref{bandassp} and 
\ref{bandassn}).

\subsection{E2 transitions}

The experimental data contain only $B({\rm E2})$ values for transitions 
corresponding in the SACM to $\Delta n_{\pi}=0$, $\Delta \lambda=0$, 
$\Delta \mu=0$, so we apply the operator (\ref{te2}) with $q_R=q_C$.  
Tables \ref{ne22e2p} and \ref{ne22e2n} show 
the transitions calculated in this approximation. 
The $q_R=q_C$ parameter was fitted to the transition with the lowest 
known experimental error \cite{bnl} and it was found to be 
1.11 $({\rm W.u.})^{1/2}$, which corresponds to 1.914 ${\rm e\ fm}^2$ for 
$^{22}$Ne. We also displayed the calculated 
results from Refs. \cite{kimura} and \cite{desc88}. We note that the 
experimental data we used \cite{bnl} differ somewhat from those displayed 
these latter works. 

The calculated values are in good agreement with the altogether six 
experimental data, two of which are lower limits and concern 
in-band transitions in the $0^+_1$ and $2^+$ bands. Note that there 
are no experimental data for transitions between negative-parity 
levels. 

The fitted parameter was also used to calculate the electric quadrupole 
moment of the $2^+_1$ state. It was found to be $-22.984$ ${\rm e\ fm}^2$, 
which is just within the error bar of the experimental value of 
$-19\pm 4$ ${\rm e\ fm}^2$ \cite{qmom}. 

\begin{table}[h]
\caption{$B({\rm E2})$ values in W.u. for transitions between states 
assigned to the same $(\lambda,\mu)$ representations. To fit the 
$q_R=q_C$ 
parameters the transition marked with asterix was used. 
Experimental data are from Ref. \cite{bnl}. 
}
\begin{tabular}{cccccccc}
$[n_{\pi}(\lambda,\mu)\kappa]_i$ & $J^{\pi}_i$ & 
$[n_{\pi}(\lambda,\mu)\kappa]_f$ & $J^{\pi}_f$ & 
$B_{\rm Exp.}({\rm E2})$ & $B_{\rm Th.}({\rm E2})$ & 
DAMD \cite{kimura} & GCM \cite{desc88} \\
\hline\\
8(8,2)0 & $2^+$ & 8(8,2)0 & $0^+$ & 12.5$\pm$0.5 & 13.38 & 14.9 & 8.4 \\
 & $4^+$ & & $2^+$ & 17.5$\pm$0.4 & 17.50$^*$ & 20.5 & 9.5 \\
 & $6^+$ & & $4^+$ & 13.7$\pm$1.7 & 16.06 & 15.8 & 10.2 \\
 & $8^+$ & & $6^+$ &  & 12.11 & 10.8 & 7.9 \\
8(8,2)2 & $3^+$ & 8(8,2)2 & $2^+$ &  & 23.88 & 15.5 & 18.5 \\
 & $4^+$ & & $2^+$ &  & 7.09 & 5.4 & 4.9 \\
 & $4^+$ & & $3^+$ &  & 17.11 & 8.4 & \\
 & $5^+$ & & $3^+$ &  & 10.59 & 7.4 & \\
 & $5^+$ & & $4^+$ &  & 12.55 & 11.4 & 6.4 \\
8(8,2)2 & $2^+$ & 8(8,2)0 & $0^+$ & $>0.26$ & 0.67 & & 0.055 \\
 & $2^+$ & & $2^+$ & $>0.21$ & 1.28 & & 0.30 \\
 & $2^+$ & & $4^+$ &  & 0.13 & & \\
 & $3^+$ & & $2^+$ &  & 1.20 & & 0.16 \\
 & $3^+$ & & $4^+$ &  & 0.97 & & \\
 & $4^+$ & & $2^+$ & $0.093\pm 0.023$ & 0.27 & & 0.70 \\
 & $4^+$ & & $4^+$ &  & 1.58 & & \\
 & $4^+$ & & $6^+$ &  & 0.44 & & \\
8(6,3)1 & $2^+$ & 8(6,3)1 & $1^+$ &  & 8.87 & & \\
 & $3^+$ & & $1^+$ &  & 10.05 & & \\
 & $3^+$ & & $2^+$ &  & 12.31 & & \\
 & $4^+$ & & $2^+$ &  & 7.80 & & \\
 & $4^+$ & & $3^+$ &  & 1.00 & & \\
 & $5^+$ & & $3^+$ &  & 11.65 & & \\
 & $5^+$ & & $4^+$ &  & 6.11 & & \\
8(4,4)0 & $2^+$ & 8(4,4)0 & $0^+$ & & 6.97 & & \\
 & $4^+$ & & $2^+$ & & 8.80 & & \\
10(14,0)0 & $2^+$ & 10(14,0)0 & $0^+$ & & 29.32 & 29.0 & 5.9 \\
 & $4^+$ & & $2^+$ & & 40.13 & 32.1 & 7.1 \\
 & $6^+$ & & $4^+$ & & 40.71 & 30.9 & 6.5 \\
 & $8^+$ & & $6^+$ & & 37.4 & 24.4 & \\
\hline\\
\end{tabular}
\label{ne22e2p}
\end{table}

\begin{table}[h]
\caption{$B({\rm E2})$ values calculated for negative-parity levels.  
}
\begin{tabular}{ccccccc}
$[n_{\pi}(\lambda,\mu)\kappa]_i$ & $J^{\pi}_i$ & 
$[n_{\pi}(\lambda,\mu)\kappa]_f$ & $J^{\pi}_f$ & 
$B_{\rm Th.}({\rm E2})$ & 
DAMD \cite{kimura} & GCM \cite{desc88} \\
\hline\\
9(11,1)1 & $2^-$ & 9(11,1)1 & $1^-$ & 36.04 & 11.4 & 15.5 \\
 & $3^-$ & & $1^-$ & 15.84 & 13.4 & 17.5 \\
 & $3^-$ & & $2^-$ & 11.44 & 12.9 & 17.2 \\
 & $4^-$ & & $2^-$ & 24.94 & 12.1 & 22.2 \\
 & $4^-$ & & $3^-$ & 11.35 & 16.8 & \\
 & $5^-$ & & $3^-$ & 22.85 & & \\
 & $5^-$ & & $3^-$ & 3.50 & & \\
9(9,2)0 & $3^-$ & 9(9,2)0 & $1^-$ & 20.08 & 41.4 & 20.6 \\
 & $5^-$ & & $3^-$ & 21.25 & 47.1 & 27.5 \\
 & $7^-$ & & $5^-$ & 18.64 & 46.8 & 17.2 \\
 & $9^-$ & & $7^-$ & 13.77 & 43.9 & \\
9(9,2)2 & $3^-$ & 9(9,2)2 & $2^-$ & 28.45 & 6.5 & \\
 & $4^-$ & & $2^-$ & 8.80 & 2.1 & \\
 & $4^-$ & & $3^-$ & 21.27 & 4.5 & \\
 & $5^-$ & & $3^-$ & 12.92 & 4.1 & \\
 & $5^-$ & & $4^-$ & 13.35 & 5.1 & \\
9(9,2)0 & $1^-$ & 9(9,2)2 & $3^-$ & 0.94 & & \\
 & $3^-$ & & $2^-$ & 1.28 & & \\
 & $3^-$ & & $3^-$ & 0.41 & & \\
 & $3^-$ & & $4^-$ & 1.32 & & \\
 & $3^-$ & & $5^-$ & 0.332 & & \\
9(7,3)1 & $2^-$ & 9(7,3)1 & $1^-$ & 26.43 & & \\
 & $3^-$ & & $1^-$ & 8.00 & & \\
 & $3^-$ & & $2^-$ & 4.04 & & \\
 & $4^-$ & & $2^-$ & 16.18 & & \\
 & $4^-$ & & $3^-$ & 9.65 & & \\
11(15,0)0 & $3^-$ & 11(15,0)0 & $1^-$ & 42.14 & & 38.8 \\
 & $5^-$ & & $3^-$ & 47.04 & & 36.5 \\
 & $7^-$ & & $5^-$ & 45.78 & & \\
 & $9^-$ & & $7^-$ & 41.20 & & \\
\hline\\
\end{tabular}
\label{ne22e2n}
\end{table}

For $q_R\ne q_C$ the selection rule concerning $(\lambda,\mu)$ is 
relaxed somewhat and transitions described in Subsection \ref{emtr} 
become allowed.

\subsection{E1 transitions}

There is a single known experimental $B({\rm E1})$ value in the 
compilation Ref. \cite{bnl}, and we used this to fit the $d_R$ 
parameter appearing in Eq. (\ref{TE1}). The results are displayed 
in Table \ref{ne22e1}. The SACM predicts relatively strong E1 
transitions from the $0^-_1$ band to the $0^+_1$ ground-state 
band and from the $2^-$ band to the $2^+$ band. Transitions from 
the $2^-$ band to the $0^+_1$ band are significantly weaker, i.e. 
they are of the order $10^{-7}$ to $10^{-6}$ W.u., while transitions 
from the $0^-_1$ to the $2^+$ band are forbidden. There are 
also transitions to the $0^+_1$ and $2^+$ bands from the bands 
assigned to the $(\lambda,\mu)=(7,3)$ SU(3) quantum numbers, 
but these are about an order of magnitude weaker than those 
from the bands labelled with (9,2). 

\begin{table}[h]
\caption{$B({\rm E1})$ values in $10^{-3}$ W.u. for transitions between 
states belonging to the $(\lambda,\mu)=(9,2)$ and (8,2) representations. 
To fit the $d_R$ 
parameter the transition marked with asterix was used. 
The experimental value for this transition is 1.5$\pm$0.4 $10^{-3}$ W.u. 
\cite{bnl}. 
}
\begin{tabular}{cccccc}
$[n_{\pi}(\lambda,\mu)\kappa]_i$ & $J^{\pi}_i$ & 
$[n_{\pi}(\lambda,\mu)\kappa]_f$ & $J^{\pi}_f$ & 
$B_{\rm Th.}({\rm E1})$ & 
GCM \cite{desc88} \\
\hline\\
9(9,2)0 & $1^-$ & 8(8,2)0 & $0^+$ & 0.83 & 0.0024 \\
 & & & $2^+$ & 1.23 & 0.49 \\
 & $3^-$ & & $2^+$ & 1.24 & 3.2 \\
 & & & $4^+$ & 0.82 & 0.91 \\
9(9,2)2 & $2^-$ & 8(8,2)2 & $2^+$ & 1.5$^*$ & \\
 & & & $3^+$ & 0.56 & \\
 & $3^-$ & & $2^+$ & 0.70 & \\
 & & & $3^+$ & 0.73 & \\
 & & & $4^+$ & 0.62 & \\
\hline\\
\end{tabular}
\label{ne22e1}
\end{table}

\section{Comparison with the results of microscopic cluster approaches}
\label{compare}

Here we compare our results with those presented in Refs. \cite{kimura} 
and \cite{desc88,desc03}. The former work \cite{kimura} employs the 
deformed-basis antisymmetrized molecular dynamics (DAMD) method and 
describes the $^{22}$Ne nucles as a $\alpha+^{16}{\rm O}$ core 
surrounded by two neutrons occupying molecular orbitals. 
In this study there are five different energy surfaces that give rise 
to six bands ($0^+_1$, $2^+$, $0^+_2$, $2^-$, $1^-$ and $0^-_1$) for 
which experimental counterparts can be identified in the spectrum of 
$^{22}$Ne. The internal structure of these bands has been described 
in Ref. \cite{kimura}, and it was found that the $1^-$, $0^+_2$ and 
$0^-_1$ bands are based on a well-developed $\alpha+^{16}{\rm O}$ 
stucture, while the $0^+_1$, $2^+$ and $2^-$ bands correspond to a 
more compact structure. Besides 
these bands $\alpha+ ^{18}{\rm O}_{\rm g.s.}$ type molecular bands are 
also discussed in Ref. \cite{kimura} by extending the calculations to 
a hybrid-GCM approach. This extension led to two more bands denoted by 
$0^+_3$ and $0^-_2$ starting near $E_x=15$ MeV. 

The other approach \cite{desc88} makes use of the generator-ccordinate 
method (GCM) and describes the $^{22}$Ne nucleus as a $^{16}$O core 
surrounded by an $\alpha$ particle and a dineutron. This method 
takes into consideration the $0^+$ ground state and the first excited 
$2^+$ state of the $^{18}$O nucleus. The internal cluster wavefunctions 
are constructed in terms of the harmonic oscillator model with a common 
oscillator parameter, and the total wavefuntion is fully antisymmetrized. 
This method gives rise to several bands both with positive and negative 
parity, furthermore, it also describes nuclear molecular type bands 
near the $\alpha+^{18}{\rm O}$ threshold. Later the GCM was also 
applied in the extended two-cluster model (ETCM), in which 
fourteen different internal $^{18}$O states were considered with 
positive parity and $J\le 4$ \cite{desc03}. 

Here we compare the band structure and the spectroscopic properties 
of $\alpha$-cluster structures in $^{22}$Ne obtained from the DAMD, 
GCM and SACM approaches. We consider separately  positive- and 
negative-parity bands at low energy, as well as nuclear molecular 
bands near the $\alpha+^{18}{\rm O}$ threshold.

\subsection{Low-lying positive-parity bands}

The members of the $0^+_1$ ground-state band are well-established 
experimentally up $J^{\pi}=8^+$ and are reproduced in all the 
models. 
In the SACM this band belongs to the (8,2) SU(3) configuration, 
which correspond to an elongated triaxial structure. The DAMD results 
\cite{kimura} also indicate a similar structure, although the 
$\alpha$+$^{16}{\rm O}$  configuration does not appear in the density 
plots. This is not a contradiction with the SACM results, because 
the model space of this latter model also includes shell-model-like 
states (SU(3) multiplets) that also appear in the cluster model space. 
The intensities of the in-band electric quadrupole transitions are also 
close in the DAMD and in the SACM, and both are rather close to the 
experimental values, as it can be seen from Table \ref{ne22e2p}. 
However, the GCM \cite{desc88} results (and the comparable ETCM 
ones \cite{desc03}) are weaker by a factor of about 0.6. 

There is some ambiguity in the assignment of experimental states to the 
$K^{\pi}=2^+$ band \cite{kimura,desc88}. Due to the lack of $B({\rm E2})$ 
data the identification of the band members is mainly based on their 
expected energy. The DAMD and SACM reproduce the band-head energy 
reasonably well, while in the GCM it comes out about 2 MeV too low. 
In all three models this band follows a regular rotational 
pattern, while the experimental energies show some odd-even staggering, 
depending on the assignment of states. 
Comparing the electric quadrupole transition intensities one finds that 
the DAMD and the GCM (as well as the ETCM) give comparable results, while 
the SACM values are somewhat larger. Transitions from this band to the 
$0^+_1$ ground-state band were calculated in the GCM and the SACM, and 
they are in general agreement both with the few experimental data and 
with each other. 

The $1^+$ band is seen in the GCM and the SACM, but not in the DAMD. 
This may be due to the relatively small deformation that may not lead 
to an easily identifiable energy surface in the latter model. 
The assignment of the lowest two states of this band is the same in 
both models, but there is a difference in the assignment of the  
states with higher $J$. Both models predict the band-head energy 
somewhat low, and both models fail to reproduce the odd-even staggering 
effect. There are no known $B({\rm E2})$ values from experiment, and 
calculations are available only in the SACM: the typical intensity of 
transitions is somewhat weaker than those within the $2^+$ band. 

The $0^+_2$ band appears in all three models, although its band head 
comes out 2 MeV too low in the GCM. In 
the SACM this band corresponds to an elongated prolate structure with 
$2\hbar\omega$ excitation, while in the DAMD it corresponds to a 
well-developed $\alpha$+ $^{16}{\rm O}$ cluster structure and two  
neutrons occupying a $pf$ orbit around it. This interpretation is 
rather similar to the SACM description. Furthermore, the $B({\rm E2})$ 
values are also rather close in the two models, 
while the GCM results fall behind them by about a factor of 5. 
It is notable that these latter values are close to those calculated for 
the (4,4)0 band in the SACM, which is expected in this region (see 
Table \ref{bandassp}) and has no correspondent in the other models. 
E2 transition from the $0^+_2$ band to the $0^+_1$ ground-state band are 
rather weak in the GCM ($B({\rm E2}) \le 0.042$ W.u. \cite{desc88}), 
and this 
is in accordance with the SACM prediction, according to which these 
transitions are forbidden in the SU(3) dynamical symmetry limit. 

Before closing this subsection it is worthwhile to compare the 
SACM, DAMD and GCM results with those obtained from the shell model 
\cite{preedom72}. In this latter work the $(sd)^6$ configuration is 
discussed, assuming an inert $^{16}$O core structure. The $T=1$ 
model space of this calculation must contain the equivalents of 
the $0\hbar\omega$ states of the SACM, i.e. the states with largest 
deformation and lowest excitation energy. Several bands are identified 
in Ref. \cite{preedom72}, including $0^+_1$ and $2^+$ seen in all three 
models discussed here. A $4^+$ band is also proposed built on the 
second $4^+$ state located at 5.52 MeV. Neither other models discussed 
here expect a $4^+$ band at such low energy, rather this state 
(with tentative $J=4$ assignment) can be interpreted as the member of 
the $2^+$ band, which exhibits a staggering pattern, and this may explain 
the relatively low energy of its $4^+$ member. A $1^+$ and a $3^+$ 
state also appear close to the experimental states at 5.332 MeV 
and 6.635 MeV, although they are not mentioned explicitly as a 
possible band. Nevertheless, a $1^+$ band is expected in this 
energy range both in the SACM and the GCM. The electric quadrupole 
transition intensities calculated in Ref. \cite{preedom72} are in 
reasonable agreement with those obtained from the other models. 
In particular, transition intensities within the $0^+_1$ band are 
rather close to the SACM and DAMD results, while 
$K^{\pi}=2^+\rightarrow 0^+_1$ 
transitions agree well with the findings of the SACM and GCM. 
Transitions within the  $2^+$ band are generally weaker than 
those predicted by either of the three models, but are close to the 
range of the DAMD and GCM results. In summary, the shell model 
results on the low-lying positive-parity states \cite{preedom72} are 
generally consistent with those obtained from the SACM, DAMD and GCM 
calculations. On the one hand this is not surprizing considering 
the shell model background of some of these models, while on the 
other hand it is remarkable that some predictions of calculations 
done 40 years ago were verified by more accurate experiments performed 
since then. 

\subsection{Low-lying negative-parity bands}

The lowest-lying negative-parity experimental band is $K^{\pi}=2^-$. In 
terms of the DAMD this is interpreted as a proton 
excited core structure with nucleon density not showing pronounced 
$\alpha$ clustering. In the SACM this band can be interpreted as 
the $(\lambda,\mu)\kappa=(9,2)2$ state with $1\hbar\omega$ 
excitation. It is notable that the nucleon density characterizing 
this band in the DAMD is rather similar to that of the $0^+_1$ 
and $2^+$ bands \cite{kimura}. This is in line with the interpretation 
within the SACM: the (8,2) and (9,2) SU(3) states have rather similar 
deformation. A $2^-$ band is also described by the GCM \cite{desc88}, 
however, it is assigned to experimental levels above 9 MeV. The 
band head also comes out close to this energy, while in the DAMD 
it is overestimated only moderately by 1.3 MeV. It seems that all 
three models have difficulty reproducing the low band-head energy. 
Theoretical $B({\rm E2})$ values are available only from the DAMD and 
the SACM calculations, but there are no experimental data to compare 
them with. 
The SACM predicts 3 to 4 times stronger transitions here (see 
Table \ref{ne22e2n}). 
Nevertheless, it is notable that the trend of the transitions 
within the band shows rather similar pattern in the two models. 
There is one experimental $B({\rm E1})$ value for a transition from 
the $2^-$ state to the $2^+$ band-head state of the $2^+$ band. 
It is relatively strong, which seems to confirm the DAMD and 
SACM expectations, according to which these two bands have similar 
structure. SACM calculations on these electric dipole transitions are 
displayed in Table \ref{ne22e1}. 

In contrast with the $2^-$ band, the $1^-$ band is interpreted 
in the DAMD as a molecular band with moderate $\alpha+^{16}$O 
cluster structure with one of the two valence neutron excited to 
a higher orbit. The same band is interpreted as a $1\hbar\omega$ 
excitation with (11,1) SU(3) character, which corresponds to  
an elongated, slightly triaxial structure. In the GCM this band 
consists of states with dominantly $\alpha+^{18}{\rm O}(2^+)$ 
character. Both the DAMD and the GCM reproduce the band-head 
energy of this band, while in the SACM it comes out about 3 MeV 
too low. 
Despite the different results on the energy spectrum, the 
electric quadrupole transitions are rather similar in the three 
models concerning both the intensity and the trend of transitions 
within this band (see Table \ref{ne22e2n}). The only exception is 
the $2^-\rightarrow 1^-$ transition, which is stronger in the SACM. 

The third negative-parity band is $0^-_1$, which is interpreted 
in the DAMD as 
a structure with prominent $\alpha+^{16}$O internal structure. 
In the SACM the lowest lying $0^-$ band is expected to be the 
one with $1\hbar\omega$ excitation and $(\lambda,\mu)\kappa=(9,2)0$ 
character. This would be the negative-parity equivalent of the 
(8,2)0 ground-state band. The band-head energy of this band comes 
out about 3 MeV too high both in the DAMD and the SACM, while it 
is only 1.2 MeV higher in the GCM, where the band assignment differs 
from that of the other models. Despite this result, the electric 
quadrupole transitions are rather similar for this band in the 
SACM and the GCM, while the DAMD results are more than a factor of 
2 stronger. As it will be discussed in the next subsection, the 
equivalent of this DAMD band might be the $0^-_2$ molecular band 
that appears both in the GCM and the SACM and has electric quadrupole 
transitions comparable to those obtained in the DAMD for the $0^-_1$ 
band. E2 transitions from the $0^-_1$ band to the $1^-$ band are 
available in the GCM and have $B({\rm E2})$ in the order of a few 
W.u., except for one transition, which is significantly stronger. 
In the SACM these transitions are zero if the $q_C=q_R$ choice is 
made in Eq. (\ref{te2}), but can acquire moderate non-zero values 
in the general case. 
Calculations are also available for electric dipole transitions to 
the ground-state band in the GCM and the SACM. These are in the order 
of $10^{-3}$ W.u. in both models, similarly to the transitions from 
the $K^{\pi}=2^-$ band to the $2^+$ band. This seems to confirm that 
these four bands may be associated to similar structure. 

It is seen that in contrast with the case of 
the positive-parity sector, the simple Hamiltonian (\ref{hami}) that 
sets the relative position of bands via the second-order SU(3) 
Casimir invariant (essentially the $Q\cdot Q$ interaction) and 
$\kappa^2$ was unable to reproduce the relative position of the 
$2^-$, $0^-_1$ and $1^-$ bands, as the first two came out too 
high, while the last one too low. The order of bands could be 
restored by a further $(-1)^{n_{\pi}+\kappa}$ type term, which 
was used in several applications of the SACM to $sd$-shell nuclei 
\cite{cunda,other}. However, in those cases the experimental spectrum 
was much richer (both in energy levels and bands), so we decided 
to omit this term in the present study in order to keep the number 
of parameters as low as possible. 

There are further negative-parity model states with $1\hbar\omega$ 
excitation predicted by the SACM, which are not seen in the DAMD and 
the GCM. These are assigned to the (7,3) 
and (5,4) SU(3) configuration and correspond to compact triaxial 
structures, similar to the (6,3) and (4,4) configurations with 
$0\hbar\omega$ excitation in the positive-parity sector of the 
spectrum. The lowest bands should start with a $3^-$ and a $4^-$ 
band-head state, and the former (7,3)3 band should be connected 
the the $2^+$ band with E1 transitions of the order $10^{-4}$ W.u. 
In Table \ref{bandassn} we tentatively assigned the (7,3)3 band-head 
state to the $3^-(7.722)$ level, however more spectroscopic data 
would be needed for a more justifiable assignment. This also applies 
to our assignment of the $J^{\pi}=4^-$ band-head state of the (5,4)4 
SACM band: since there are no $4^-$ states in the experimental 
energy spectrum \cite{bnl}, we tentatively chose the $(3)^-$ state 
at $E_x=8.376$ MeV (see Table \ref{bandassn}).

\subsection{Molecular orbital bands}

All three models predict both positive- and negative-parity nuclear 
molecular bands with large deformation and pronounced $\alpha$-cluster 
character a few MeV above the $\alpha+^{18}{\rm O}$ threshold. In 
the DAMD these bands are obtained from a hybrid-GCM extension by 
including $\alpha$+$^{18}{\rm O}_{\rm g.s.}$ configuration in the 
basis \cite{kimura}. The extended model predicts a $0^+$ 
and a $0^-$ band starting near 14-15 MeV. In the GCM approach 
\cite{desc88} there appears the $0^+_3$ band starting at 10.28 MeV 
and the $0^-_2$ band starting at 11.47 MeV. The $0^+_3$ band has 
dominantly $\alpha+^{18}{\rm O}(0^+)$ character for low $J$, while 
for increasing $J$ the $\alpha+^{18}{\rm O}(2^+)$ character 
increases. This latter structure dominates the $0^-_2$ band. 
Similar $\alpha$-cluster bands are also present in the SACM basis 
in this energy range. A difference with respect to the 
DAMD calculations is that the $\alpha$+$^{18}{\rm O}$ 
model space of the SACM also contains configurations with the 
first excited $2^+$ (and $4^+$) state of $^{18}{\rm O}$. Actually, the 
model states contain a mixture of these states and the ground state 
of $^{18}{\rm O}$ in proportions prescribed by the SU(3) symmetry. 
This mixed character makes them similar to the states obtained 
in the GCM calculations. 

Furthermore, besides the $0^+$ and $0^-$ bands with this 
structure the SACM also predicts $1^+$, $2^+$, $1^-$ and $2^-$ bands 
with the same $n\hbar\omega$ excitation. These latter bands have a 
slightly triaxial structure, while the $(n,0)0$ type bands have 
definite prolate character. Bands with $K\ne 0$ contain also 
unnatural-parity levels, but the experimental identification of 
such levels may be difficult. 
According to Fig. \ref{ne22bh} the best candidates for the 
positive-parity molecular orbital bands are the $2\hbar\omega$ 
(12,1)1, (10,2)0 and the $4\hbar\omega$ (16,0)0 states of the SACM. 
The simple Hamiltonian (\ref{hami}) with parameters fitted mainly 
to the lower-lying levels gives as band-head energy 12.067, 15.803 and 
18.758 MeV, respectively. The in-band E2 transitions are rather 
strong (larger than 25 W.u.). These values are close to those obtained in 
the GCM for the $0^+_3$ band (40 to 50 W.u.) \cite{desc88}. In the 
GCM there are also weak E2 transitions from this band to the $0^+_1$ 
ground-state band (0.7 to 2.8 W.u.). In the SU(3) dynamical symmetry 
approximation of the SACM the corresponding transitions are forbidden 
from the (12,1)1 and (16,0)0 bands, but they are allowed from the 
(10,2)0 band. Calculations on the electromagnetic transition data 
are not available for the $0^+_3$ band in the DAMD \cite{kimura}. 
The deformation of this band that is characterized by the rotational 
constant is also close in the DAMD and the GCM, and is in the range 
obtained for the (12,1)1 and (10,2)0 SACM bands. 

There is another positive-parity band in the ETCM \cite{desc03} 
expected at $E_x$=12.8 MeV and identified as $0^+_5$. This band is 
also characterized by strong ($\sim 45$ W.u.) E2 transitions and very 
weak transitions ($<1$ W.u.) to the ground-state band. The rotational 
constant of this band is similar to that of the $0^+_3$ band discussed 
above, which, however, does not appear in the ETCM calculations. 

The $0^-_2$ negative-parity molecular orbital band is seen in the GCM 
\cite{desc88} and the hybrid-GCM extension of the DAMD \cite{kimura} at 
11.47 and 14.8 MeV, respectively. The typical in-band E2 transitions 
are found rather strong (35 to 40 W.u.) in the GCM, while there is no 
similar calculation in Ref. \cite{kimura}. Candidates for this band in 
the SACM can be those with the SU(3) labels (15,0)0 or (13,1)1 starting 
at 12.8 and 18.1 MeV, respectively. Transitions in both come out in the 
range of $<45$ W.u. It is notable that in the DAMD similar $B({\rm E2})$ 
values are obtained for the $0^-_1$ band, which has the most prominent 
$\alpha$-cluster character in the DAMD according to Ref. \cite{kimura}. 
It is also suggested that this and the $0^+_2$ band are parity doublets. 
Since in the SACM the latter band is assigned to (14,0)0, a natural 
interpretation of this finding would be assigning (15,0)0 to the 
$0^-_1$ band of the DAMD. 

The ETCM \cite{desc03} also describes a negative-parity band at 
$E_x=12.58$ MeV, which is expected to contain the close-lying pairs 
of $1^-$, $3^-$, $7^-$ and $9^-$ states found experimentally 
\cite{rogachev}. However, the calculated in-band electric quadrupole 
transitions are much weaker ($<15$ W.u.) than those found for the 
molecular orbital bands in the GCM and the SACM. Furthermore, the 
ETCM predicts relatively strong ($10^{-3}$ W.u.) electric dipole 
transitions to the ground-state band indicating that they have similar 
structure. In the SACM these features are similar to those of the (7,3)1 
band, which is expected to appear in this energy domain. The splitting 
of $\alpha$-cluster states \cite{rogachev} is reproduced only in the 
ETCM, which associates rather rich structure to the core $^{18}$O 
nucleus.

\section{Summary and conclusions}
\label{concl}

There are several bands established experimentally in the spectrum 
of $^{22}$Ne (e.g. $0^+_1$, $2^+$, $2^-$), however, the assignment of 
levels to further bands is generally ambiguous due to the lack of 
spectroscopic data. The SACM describes all the well-established bands 
rather well especially in the positive-parity sector, while the band-head 
energies are less well reproduced for negative parity. This seems to 
indicate that the simple five-parameter phenomenological Hamiltonian of 
the SACM may be a crude approximation, at least for negative-parity 
bands. It is notable though, that some band-head energies are 
not reproduced accurately in other models either. 

The comparison with the results of microscopic models such as the 
DAMD \cite{kimura} and the GCM \cite{desc88} reveals that there are 
many similarities with the predictions of these models. The three 
models describe bands with relatively large deformation both for positive 
and negative parity and the expected properties of these bands ($0^+_1$, 
$2^+$, $0^+_2$, $1^-$, $0^-_2$) generally coincide. The $1^+$ band is 
seen only in the SACM and the GCM but not in the DAMD, which may 
indicate that the corresponding energy surface might not be identified 
in the latter model due to the small deformation. 
It is notable that the results of the three cluster models on low-lying 
positive-parity states agree rather well with shell model calculations 
\cite{preedom72} assuming an $(sd)^6$ configuration and an inert $^{16}$O 
core. In the case of the SACM this may be explained by the fact that 
the SACM model space with $0\hbar\omega$ excitation is a subset of the 
harmonic oscillator shell model. 

The $0^+_2$ band has well-developed $\alpha$-cluster structure with 
large deformation in the SACM and the DAMD, while these features are 
less pronounced in the GCM. In this latter model the $0^+_3$ band 
has marked $\alpha$-cluster structure, together with the $0^-_2$ 
band. Such molecular bands appear both in the SACM and the DAMD, 
although in the latter case a hybrid $\alpha+^{18}{\rm O}_{\rm g.s.}$ 
extension of the model is necessary. 

The internal excitation of the $^{18}$O cluster plays an important 
role both in the GCM and the SACM. In the GCM several bands are interpreted 
as structures built on the $2^+_1$ level of $^{18}$O, while this 
configuration and that with the $4^+_1$ state appear in the SACM 
states due to the assumed SU(3) symmetry. There are relatively numerous 
calculated values on electric quadrupole and dipole transitions in the 
GCM, which allows a straightforward comparison of the two models. The 
selection rules of these transitions are rather similar in the two 
models in the sense that transitions forbidden due to the assumed 
SU(3) dynamical symmetry in the SACM correspond to rather weak transitions 
in the GCM. These results indicate that the $0^+_1$, $2^+$ and $2^-$ 
bands, as well as the $0^-_1$ band present in both models may have 
similar structure. This feature is also confirmed by the DAMD calculations 
with the exception that the $0^-_1$ band there has more pronounced 
deformation and $\alpha$-cluster structure. It is possible that this 
DAMD band should correspond to the $0^-_2$ band of the GCM and the 
(15,0) SU(3) band in the SACM. 

The experimental confirmation of the predictions given by the three 
models is missing in most cases, especially concering electric quadrupole 
and dipole transitions. Such spectroscopic information would also make it 
possible to put the tentative band assignments on a more firm basis. 

The similariries between the band structure and the properties of bands 
described simultaneously in the SACM and GCM have already been noticed 
in earlier calculations on $\alpha$-cluster states of other nuclei, 
such as $^{18}$O \cite{prc92}. It is probable that this similarity 
originates from the fact that the GCM employs an harmonic oscillator 
basis with a common oscillator constant for each cluster. The SU(3) 
dynamical symmetry of the SACM is also based on this harmonic oscillator 
assumption. The fully microscopic model space of this model contains 
SU(3) shell model basis constructed from antisymmetrized harmonic 
oscilator states. This common background may be manifested in similar 
matrix elements and selection rules despite the different construction 
of the operators. 

A systematic study of core+$\alpha$ states of nuclei in the $^{20}$Ne 
to $^{24}$Mg region seems worthwhile. A similar study has already been 
performed in the region between the $^{16}$O to $^{20}$Ne nuclei and 
it was found that the SACM parameters vary smoothly in this domain 
\cite{plb96}. 

\begin{acknowledgments}
This work was supported by the OTKA (grant no. K72357) and the JSPS--MTA 
Japanese--Hungarian cooperation. The author is indebted to M. Kimura 
for valuable discussions.
\end{acknowledgments}

\end{document}